\documentclass[prb,superscriptaddress,twocolumn,amsmath,amssymb,10pt,nofootinbib]{revtex4}
\usepackage{graphicx} 
\input{ObrienReviewHeader}
\setboolean{ispreprint}{false}
\setboolean{versioncontrol}{false}

\ifthenelse{\boolean{ispreprint}}{
}
{
    \textheight=9.5in
    \textwidth=7.5in
    \columnwidth=3.5in
    \oddsidemargin=-0.5in
}

\begin{document}
\title{Quantum Computing}
\date{June 15, 2009}

\author{T. D. Ladd}
\affiliation{Edward L. Ginzton Laboratory, Stanford University,
Stanford, California 94305-4088, USA}

\author{F. Jelezko}
\affiliation{Physikalisches Institut, Universit\"{a}t
Stuttgart, Pfaffenwaldring 57, D-70550, Germany}

\author{R. Laflamme}
\affiliation{Institute for Quantum Computing and Department of
Physics and Astronomy, University of Waterloo, 200 University
Avenue West, Waterloo, ON, N2L 3G1, Canada}
\affiliation{Perimeter Institute, 31 Caroline Street North,
Waterloo, ON, N2L 2Y5, Canada }

\author{Y. Nakamura}
\affiliation{Nano Electronics Research Laboratories, NEC
Corporation, Tsukuba, Ibaraki 305-8501, Japan}
\affiliation{Frontier Research System, The Institute of
Physical and Chemical Research (RIKEN), Wako, Saitama 351-0198,
Japan}

\author{C. Monroe}
\affiliation{Joint Quantum Institute, University of Maryland Department of Physics \\
    and National Institute of Standards and Technology, College Park, MD 20742, USA}

\author{J. L. O'Brien}
\affiliation{Centre for Quantum Photonics, H. H. Wills Physics
Laboratory \& Department of Electrical and Electronic
Engineering, University of Bristol, Merchant Venturers
Building, Woodland Road, Bristol, BS8 1UB, UK}

\begin{abstract}
\version{Abstract version 1.1, written by Jeremy, slightly
revised by Thaddeus, added 1/6/09}

Quantum mechanics---the theory describing the fundamental
workings of nature---is famously counterintuitive: it predicts
that a particle can be in two places at the same time, and that
two remote particles can be inextricably and instantaneously
linked. These predictions have been the topic of intense
metaphysical debate ever since the theory's inception early
last century. However, supreme predictive power combined with
direct experimental observation of some of these unusual
phenomena leave little doubt as to its fundamental correctness.
In fact, without quantum mechanics we could not explain the
workings of a laser, nor indeed how a fridge magnet operates.
Over the last several decades quantum information science has
emerged to seek answers to the question: can we gain some
advantage by storing, transmitting and processing information
encoded in systems that exhibit these unique quantum
properties? Today it is understood that the answer is yes. Many
research groups around the world are working towards one of the
most ambitious goals humankind has ever embarked upon: a
quantum computer that promises to exponentially improve
computational power for particular tasks. A number of physical
systems, spanning much of modern physics, are being developed
for this task---ranging from single particles of light to superconducting circuits---and it is not yet clear
which, if any, will ultimately prove successful. Here we
describe the latest developments for each of the leading
approaches and explain what the major challenges are for the
future.

\end{abstract}
\maketitle

\section{Introduction}
\version{Version 1.0, written by Jeremy, added 1/6/09\\
    Version 2.0, written by Thaddeus, added 1/8/09,\\
    Revision 2.1, shortened to include/transition to resource section.\\
    Awaiting comments from other authors!}

One of the most bizarre and fascinating predictions of the
theory of quantum mechanics is that the information processing
capability of the universe is much larger than it seems.  As
the theory goes, a collection of quantum objects inside a
closed box will in general proceed to do everything they are
physically capable of, all at the same time. This closed system
is described by a \qu{wave function}, which for more than a few
particles is an incredibly large mathematical entity describing
states of matter and energy far beyond experience and
intuition. The wave function, however, is only maintained until
the box is opened and the system \qu{collapses} randomly into
one particular \qu{classical} outcome. Erwin Schr\"odinger
attempted to reduce these notions to absurdity by connecting
the known quantum behavior of an atomic nucleus to a cat in a
box that becomes simultaneously alive and dead before the box
is opened. Schr\"odinger intended for the difficulty of
imagining a cat in a \qu{superposition} of alive and dead to
make us question whether this quantum theory could possibly be
correct.

And yet, nearly a century later, quantum theory has yet to fail
in predicting an experiment.  Although observing an actual
\qu{alive and dead} cat is still beyond experimental
capabilities, a number of useful technologies have arisen from
the counterintuitive quantum world.  The quantum computer, a
device which uses the full complexity of a many-particle
wavefunction to solve a computational problem, may soon be one
of these technologies.

The nature and purpose of quantum computation are often
misunderstood.  The context for the development of quantum
computers may be clarified by comparison to a more familiar
quantum technology: the laser. Before the invention of the
laser we had the sun, and fire, and the lantern, and then the
lightbulb. Despite these advances in making light, until the
laser this light was always \qu{incoherent}, meaning that the
many electromagnetic waves generated by the source were emitted
at completely random times with respect to each other.  One
possibility allowed by quantum mechanics, however, is for these
waves to be generated in phase, and by engineering and
ingenuity methods were discovered for doing so, and hence came
about the laser.  But lasers do not replace light bulbs for
most applications; instead, they produce a different kind of
light---coherent light---which is useful for thousands of
applications from eye surgery to cat toys, most of which were
unimagined by the first laser physicists.

Likewise, a quantum computer will not necessarily be faster,
bigger, or smaller than an ordinary computer.  Rather, it will
be a different kind of computer, engineered to control
coherent quantum mechanical waves for different
applications.  The result will be a \qu{closed box}, designed to
simultaneously perform everything it is physically capable of,
all at once, with all of those possibilities focused toward a
computational problem whose solution will be observable after
the box is opened.

So what will be in the box, and what will it be able to do?
Both questions are currently subjects of ongoing research.  The
first question will be addressed in ensuing sections; the
second is worthy of a review of comparable size, and interested
readers are advised to see Ref.~\onlinecite{ncbook}.  For now,
we provide only a brief synopsis of quantum computer
\qu{software}.


One example of a task for a quantum computer is the quantum
fourier transform, which continues the exponential increase in
computational efficiency begun by the fast fourier
transform\cite{ct65}. This subroutine is at the core of Peter
Shor's seminal quantum algorithm for factoring large
numbers\cite{shor94}, which is one among several quantum
algorithms that would allow modestly sized quantum computers to
outperform the largest classical supercomputers in solving the
specific problems required for decrypting encoded information.
Although these algorithms have done much to spur the
development of quantum computers, another application is likely
to be far more important in the long term. This application is
the first envisioned for quantum computers, by Richard Feynman
in the early 1980s\cite{feynman82}: the efficient simulation of
that large quantum universe underlying all matter.  Such
simulations may seem to lie in the esoteric domain of research
physics, but these same quantum laws govern the behavior of the
many emerging forms of nanotechnology, including nature's
nanomachinery of biological molecules.   The engineering of the
ultra-small will continue to advance and change our world in
coming decades, and as this happens we will likely use quantum
computers to understand and engineer such technology at the
atomic level.

Quantum information research promises more than computers, as
well. Similar technology allows quantum communication, which
enables the sharing of secrets with security guaranteed by the
laws of physics. It also allows quantum metrology, in which
distance and time are measured with higher precision than would
be possible otherwise. The full gamut of potential technologies
have probably not yet been imagined, nor will it be until
actual quantum information hardware is available for future
generations of quantum engineers.

This brings us to the central question of this review: what
form will quantum hardware take?  Here there are no easy
answers.
Quantum computers are often imagined to be constructed by controlling the smallest form of matter, isolated atoms, as in ion traps and optical lattices, but they may likewise be made from electrical
components far larger than routine electronic components, as in
superconducting phase qubits, or even from a vial of liquid, as
in Nuclear Magnetic Resonance (\NMR).  Of course it would be
convenient if a quantum computer can be made out of the same
material that current computers are made out of, i.e. silicon,
but it may be that they will be made out of some other material
entirely, such as InAs quantum dots or microchips made of
diamond.

In fact, very little ties together the different
implementations of quantum computers currently under
consideration.
We provide a few general statements about requirements in the
next section, and then describe the diverse technological
approaches for satisfying these requirements.

\section{Requirements for Quantum Computing}
\label{sec:resources}
\version{Version 1.0, written by Thaddeus (including comments),
added 3/9/09\\
Version 1.1, revised by Thaddeus, converted to a section
because of its length, stealing text from introduction\\
Version 1.2 minor changes by Thaddeus, some references added 3/31/09\\
waiting for comments!}
\label{commonbox}

Perhaps the most critical, universal aspect of the various
implementations of quantum computers is the \qu{closed box}
requirement: a quantum computer's internal operation, while
under the programmer's control, must otherwise be out of
contact with the rest of the universe.  Small amounts of
information-exchange into and out of the box can disturb the
fragile, quantum mechanical waves that the quantum computer
depends on, causing the quantum mechanically destructive
process known as decoherence, discussed further in Sec.~\ref{sec:decoherence}.
Unfortunately no system is fully free of decoherence, but a
critical development in quantum computer theory is the ability
to correct for small amounts of it through various techniques
under the name of Quantum Error Correction (\QEC). In \QEC,
entropy introduced from the outside world is flushed from the
computer through the discrete processes of measuring and
re-initializing qubits, much as digital information today
protects against the noise sources problematic to analog
technology.  Of course, the correction of errors may be useless
if the act of correcting them creates more errors. The ability
to correct errors using error-prone resources is called
fault-tolerance\cite{shor96}. Fault-tolerance has been shown to
be theoretically possible for error rates beneath a critical
threshold that depends on the computer hardware, the sources of
error, and the protocols used for \QEC.  Realistically, most of
the resources a fault-tolerant quantum computer will use will
be in place to correct its own errors.  If computational
resources are unconstrained, the fault-tolerant threshold can
be as high as 3\%\cite{knill05}.

An early characterization of the physical requirements for an
implementation of a fault-tolerant quantum computer was carried out by
David DiVincenzo\cite{divincenzo00}.  However, since that time the
ideas for implementing quantum computing have diversified, and
the DiVincenzo criteria as originally stated are difficult to
apply to many emerging concepts.
Here, we rephrase DiVincenzo's original considerations into
three, more abstract criteria, and in so doing introduce a
number of critical concepts common to most quantum technologies.\\

\noindent\textit{\bf 1. Scalability: the computer must operate
in a Hilbert space whose dimensions may be grown exponentially
without an exponential cost in resources (such as time, space or energy.}

The standard way to achieve this follows the first DiVincenzo
criterion: one may simply add well-characterized \emph{qubits}
to a system.  A qubit is a quantum system with two states,
$\ket{0}$ and $\ket{1}$, such as a quantum spin with $S=1/2$.
The logic space available on a quantum system of $N$ qubits is
described by a very large group [known as SU($2^N$)], which is
much larger than the comparable group [SU(2)$^{\otimes N}$] for
$N$ unentangled spins or for $N$ classical bits. Ultimately, it is this large space that provides a quantum computer its power.  For qubits, the size and energy of a quantum computer generally grows linearly with $N$.

Although qubits are a convenient way to envision a quantum
computer, they are not a prerequisite.  One could use quantum
$d$-state systems (qudits) instead, or even the continuous
degrees of freedom available in laser-light.  In all cases,
however, an exponentially large space of accessible quantum
states must be available.


In principle, there is an exponentially large Hilbert space in
the bound states a single hydrogen atom, a system which is
clearly bounded by the Rydberg energy of 13.6 eV and consists
of only two particles! However, the states of a hydrogen atom
in any realistic experiment have a finite width due to
decoherence, limiting the useful Hilbert space (for which
DiVincenzo introduced his third criterion; see Sec.~\ref{sec:decoherence}). Further,
access to an exponentially large set of a hydrogen atom's
states comes at the exponentially large cost in the size of
that atom and the time required to excite it to any arbitrary
state\cite{b-kcd02}.

While it is straightforward to see why a single-atom quantum
computer is \qu{unscalable},  declaring a technology \qu{scalable} is
a tricky business, since the resources used to define and
control a qubit are diverse.  They may include space on a
microchip, classical microwave electronics, dedicated lasers,
cryogenic refrigerators, etc. For a system to be scalable,
these \qu{classical} resources must be made scalable as well,
which tie into complex engineering issues and the
infrastructure available for large-scale
technologies.\\

\noindent{\it\bf 2. Universal Logic: the large Hilbert space
must be accessible using a finite set of control operations; the resources for this set must also not grow exponentially.}

In the most standard picture of computing, this criterion
(DiVincenzo's fourth) means that a system must have available a
universal set of quantum logic gates. In the case of qubits, it
is sufficient to have available any \qu{analog} single-qubit gate
(e.g. an arbitrary rotation of a spin-qubit), and almost any
\qu{digital} two-qubit logic operation, such as the
\CNOT\ gate.

But quantum computers need not be made with gates. In
\emph{adiabatic quantum computation}\cite{adiabatic_first}, one defines the answer to a computational problem as the ground state of a complex
network of interactions between qubits, and then one
adiabatically evolves those qubits into that ground state by
slowly turning on the interactions. In this case, evaluation of
this second criterion requires that one must ask whether the
available set of interactions is complex enough, how long it
takes to turn on those interactions, and how cold the system
must be maintained. As another example, in \emph{cluster-state
quantum computation}\cite{rb01}, one particular quantum state (the cluster state) is generated in the computer through a very small set of non-universal quantum gates, and then computation is performed
by changing the way in which the resulting wave function is
measured. Here, the measurements provide the \qu{analog}
component that completes the universal logic. Adiabatic and
cluster-state quantum computers are provably equivalent in
power to gate-based quantum computers\cite{adiabatic_simple_proof}, but their implementation may be simpler for some technologies.

One theoretical issue in the design of fault-tolerant quantum
computers is that for most \QEC\ protocols, \qu{digital}
quantum gates (or, more precisely, those in the Clifford group)
are relatively easy to perform fault-tolerantly on encoded
qubits, while the \qu{analog} (non-Clifford) quantum gates are
substantially more challenging.  In other protocols, the analog
gates may become easy, and then the digital ones become
difficult.  The modern design of fault-tolerant protocols
centers around maintaining universality and balancing the
difficulties between the two types of operations.

No matter what scheme is used, however, \QEC\ fundamentally
requires the third abstract criterion:\\

\noindent{\it\bf 3. Correctability: It must be possible to
extract the entropy of the computer to maintain the computer's
quantum state.}

Regardless of \QEC\ protocol, this will require some
combination of efficient \emph{initialization} (DiVincenzo's
second criterion) and \emph{measurement} (DiVincenzo's fifth
criterion). \emph{Initialization} refers to the ability to
quickly cool a quantum system into a low-entropy state; for
example, the polarization of a spin into its ground state.
\emph{Measurement} refers to the ability to quickly determine
the state of a quantum system with the accuracy allowed by
quantum mechanics.  It is possible that these two abilities are
the same.  For example, a \emph{quantum non-demolition} (\QND)
measurement alters the quantum state by projecting to the measured state, which remains the same even after repeated measurements.  Clearly, performing a \QND\ measurement also initializes the quantum
system into the state measured.  Some \QND\ measurements also
allow quantum logic; they are therefore quite powerful for
quantum computing. The relationship between the need for
initialization and measurement is complex; depending on the
scheme used for fault-tolerance, one may generally be replaced
by the other.  Of course, some form of measurement is always
needed to read out the state of the computer at the end of a
computation.  Notably, the amount of required physical
initialization is not obvious, as schemes have been developed
to quantum compute with states of high entropy\cite{kl98}.

Quantum computation is difficult because the three basic
criteria we have discussed appear to be conflicted.  For
example, those parts of the system in place to achieve rapid
measurement must be turned strongly \qu{on} for error
correction and read-out, but must be turned strongly \qu{off}
to preserve the coherences in the large Hilbert space.
Generally, neither the \qu{on} state nor the \qu{off} state are
as difficult to implement as the ability to switch between the
two!

DiVincenzo introduced extra criteria related to the ability to
communicate quantum information between distant qubits, for
example by converting stationary qubits to \qu{flying qubits}
such as photons.  This ability is important for other
applications of quantum processors such as quantum repeaters\cite{dur}, but the ability to add non-local quantum communication also
substantially aids the scalability of a quantum computer
technology.  Quantum communication allows small quantum
computers to be \qu{wired together} to make larger ones, it
allows specialized measurement hardware to be located distant
from sensitive quantum memories, and it makes it easier to
achieve the strong qubit-connectivity required by most schemes
for fault-tolerance.

Evaluating the resources required to make a quantum technology
truly scalable is an emerging field of quantum computer
research, known as quantum computer architecture.  Successful
development of quantum computers will require not only further
hardware development, but also the continued theoretical
development of algorithms and \QEC, and the architecture
connections between the theory and the hardware.  These efforts strive to find ways to maintain the simultaneous abilities to control quantum systems, to measure them, and to preserve their strong isolation from uncontrolled parts of their environment.   The simultaneity of these aspects forms the central challenge in actually building quantum computers, and in the ensuing sections, we introduce the various technologies researchers are currently employing to solve this challenge.

\section{Quantifying Noise in Quantum Systems}
\label{sec:decoherence}
\version{Version 0.5, written by Thaddeus added 3/9/09\\
Version 0.6, added to benchmarking table from Ray, 3/17/09\\
Version 0.9, Thaddeus added decoherence table (unfinished), 3/27/09\\
Version 1.0, Table suggestions added by Yasu, 4/3/09\\
Version 1.1, References from Ray added}

A key challenge in quantum computation is handling noise.  For
a single qubit, noise processes lead to two types of
relaxation.  First, the energy of a qubit may be changed by its
environment in a random way which, on-average, brings the qubit
to thermal equilibrium with its environment.  The timescale for
this equilibration is $T_1$.  Typically, systems used for
qubits have long $T_1$ timescales, which means that $T_1$ can
usually be ignored as a computation error.  However, in many
experimental systems, $T_1$ sets the timescale for
initialization.

More dangerous for quantum computing are processes which
randomly change the phase of a qubit; i.e. processes that
scatter a superposition such as $\ket{0}+\ket{1}$ into
$\ket{0}+\exp(i\phi)\ket{1}$, for an unknown value of $\phi$.
This is known as decoherence, and the timescale for phase
randomization by decoherence is called $T_2.$ The processes
leading to $T_1$ also contribute to $T_2$, resulting in $T_2$
being upper bounded by $2T_1$. But $T_2$ processes cost no
energy, and as a result may be much more frequent than $T_1$
processes.

In studying noise, one must average over a large ensemble of
measurements.  It is frequently the case that in this ensemble
of measurements, the energy of a qubit is slightly different in
each measurement.  As a result, superpositions again develop
unknown phases, and as a result effects appear which resemble
those contributing to $T_2$.  This process is known as
dephasing, and it occurs on a timescale $T_2^* \le T_2$.
However, the phase evolution that contributes to $T_2^*$ is
constant for each member of the ensemble, and may therefore be
reversed. The standard method for doing so is known as the
spin-echo, following the \NMR\ technique developed in
1950\cite{hahn1950}. By unconditionally flipping the state of a
qubit after a time $\tau$, and then allowing evolution for
another time $\tau$, any static phase evolution is reversed,
leading to an apparent ``rephasing." Through spin-echo
techniques, the effects of decoherence ($T_2$) can be
distinguished from those of dephasing ($T_2^*$).

The value of $T_2$ is used as an initial characterization of
many qubits, since, at a bare minimum, qubits need to be
operated much faster than $T_2$ to allow fault-tolerant quantum
computation.  This is the third DiVincenzo criterion.
However, $T_2$ is \emph{not} the timescale in which an entire
computation takes place, since \QEC\ may correct for phase
errors.  Also, the measured values of $T_2$ are not fundamental
to a material and a technology.  Generally, $T_2$ can be
extended by a variety of means, such as defining qubits with
\emph{decoherence free subspaces}\cite{lcw98} which are less
sensitive to noise; applying \emph{dynamic decoupling
techniques}\cite{CarrPurcell,CPMG,lv98,udd,magical_udd,universal_udd},
such as the spin-echo itself, to periodically reverse the
effects of environmental noise; or simply improving those
aspects of the apparatus or material that leads\ to the $T_2$
noise process in the first place.

Other noise processes exist besides $T_1$ and $T_2$ relaxation.
Large-dimensional systems, such as multiple-coupled qubits, may
be hurt by noise processes distinct from single-qubit $T_1$ and
$T_2$ processes. Also, some qubits suffer noise processes that
effectively remove the qubit from the computer, such as loss of
a photon in a photonic computer or the scattering of an atom
into a state other than a qubit state.  These processes may
also be handled by error correction techniques.

In practice, once relaxation times are long enough to allow
fault-tolerant operation, imperfections in the coherent control
of qubits are more likely to limit a computer's performance. As
devices are scaled up to a dozen of qubits, the use of state
and process tomography, useful to fully understand the
evolution of very small quantum systems, becomes impractical.
For this reason, protocols that assess the quality of control
in larger quantum processors have been developed.
These enable a characterisation of gate fidelity that can be
used to benchmark various technologies.

The table below gives measured $T_2$ decoherence times and the
results of one-qubit and multi-qubit benchmarking or tomography
for several technologies.

\ifthenelse{\boolean{ispreprint}}
{
\newpage
\addtolength{\tabcolsep}{2pt}
\newcommand{\thebigtablecaption}{
Table comparing the current performance of various matter
qubits.  The approximate resonant frequency of each qubit is
listed as $\omega_0/2\pi$; this is not necessarily the speed of
operation, but sets a limit for defining the phase of a single
qubit.
%
Therefore, $Q=\omega_0T_2$ is a very rough quality factor.
Benchmarking values show approximate error rates for single or
multi-qubit gates. Values marked with * are found by state
tomography, and give the departure of the fidelity from 100\%.
Values marked with \dag\ are found with randomized
benchmarking.  Other values are rough experimental gate error
estimates.}
%
%
    \begin{tabular}[c]{|rl|rrl|c|c|}
      \hline
      & \multirow{2}{2in}{\centering Type of Matter Qubit\mystrut}  &
      \multicolumn{3}{c|}{Coherence\mystrut} &
      \multicolumn{2}{c|}{Benchmarking\mystrut}
      \\\cline{3-7}\mystrut
      &
      & $\omega_0/2\pi$
      & \multicolumn{1}{c}{$T_2$}
      & $Q$
      & \multicolumn{1}{c|}{1 qbit}
      & \multicolumn{1}{c|}{2 qbit}
	\\
      \hline
      \multirow{3}{2.5ex}{\rotatebox{90}{\makebox[9ex][c]{\abbrev{AMO}~\mystrut}}}
		&
        \mystrut
        Trapped Optical Ion\cite{optical_ion_coherence,iontrap_multi} ($^{40}$Ca$^{+}$) 
        & 400\THz
        & 1\ms
        & $10^{12}$
        & $0.1\%$
        & $0.7\%^*$ 
        \\
        & Trapped Microwave Ion\cite{Langer_coherence,iontrap_single,NIST_geometric} ($^{9}$Be$^+$)
        & 300\MHz
        & $10\sec$
        & $10^{10}$
        & $0.48\%^\dag$ 
        & 3\%
        \\
        & Trapped Neutral Atoms\cite{microtrap_coherence} ($^{87}$Rb)
        & 7\GHz
        & $3\sec$
        & $10^{11}$
        & 5\%
        &
        \\
        & Liquid Molecule Nuclear Spins\cite{NMR}
        & 500\MHz
        & 2\seconds
        & $10^9$
        & $0.01\%^\dag$  
        & $0.47\%^\dag$ 
        \\
    \hline
    \multirow{8}{2.5ex}{\rotatebox{90}{{Solid-State~\mystrut}}}
        & \mystrut
	   e$^{-}$ Spin in GaAs Quantum Dot\cite{pjtlylmhg05,bayerT2,pressnature}
	 & 10\GHz
	 & 3\us
	 & $10^5$
        & 5\%
        &
\\	
        & e$^{-}$ Spins Bound to $^{31}$P:$^{28}$Si\cite{tlar03,mtbslashal08}
        & {10\GHz}
        & {60\ms}
        & {$10^9$}
        & 5\%
        & 10\%
\\
        & Nuclear Spins in Si\cite{lmyai05}
        & 60\MHz
        & 25~sec
        & $10^9$
        & 5\%
        &
\\
        & NV$^-$ Center in Diamond\cite{diamond_ultralongT2,jelezko_04a,diamond_entanglement}
        & 3\GHz
        & 2\ms
        & $10^7$
        & 2\%
        & 5\%
\\
        & Superconducting Phase Qubit\cite{Neeley08,lucero08,steffen06}  
        & 10\GHz
        & $350\ns$ 
        & $10^4$
        & $2\%^*$  
        & $24\%^*$
	 \\
        & Superconducting Charge Qubit\cite{schreier08,chow09,DiCarlo09}  
        & 10\GHz
        & 2\us
        & $10^5$
        & $1.1\%^\dag$ 
        & $10\%^*$ 
      \\
        & Superconducting Flux Qubit\cite{Bertet05,plantenberg07}
        & 10\GHz
        & 4\us
        & $10^5$
        & 3\%
        & 60\%
\\
\hline
    \end{tabular}
\ifthenelse{\boolean{ispreprint}}
{
\\
\thebigtablecaption } {
\parbox{1.9in}{
\textsf{\small\raggedright\thebigtablecaption}} }

}
{
\begin{widetext}

\end{widetext}
}

\section{Cavity Quantum Electrodynamics}
\label{sec:cqed}
\version{Version 1.0, written by Jeremy, added 1/6/09\\
         Version 1.5, modified by Thaddeus, 3/22/09}

Many concepts for scalable quantum computer architectures
involve wiring distant qubits via communication using the
electromagnetic field, e.g. infrared photons in fiber-optic
waveguides or microwave photons in superconducting transmission
lines. Unfortunately, the interaction between a single qubit
and the electromagnetic field is generally very weak.  For
applications such as measurement, in which quantum coherence is
deliberately discarded, using more and more photons in the
electromagnetic field can sometimes be enough.  However,
photons easily get lost into the environment, which causes
decoherence, and this happens more quickly with stronger
fields. Coherent operation requires coupling qubits to weak,
single-photon fields with very low optical loss.  Such coupling
becomes available when discrete, atom-like systems are placed
between mirrors that form a high-quality cavity, introducing
the physics known as cavity quantum electrodynamics
(\CQED)\cite{mabuchiCQED}.  Cavity \QED\ has been an important
topic of fundamental research for many
years\cite{kimble_early,tu-prl-75-4710,no-nat-400-239,ye-prl-83-4987},
and was employed for one of the earliest proposals for quantum
computing\cite{pe-prl-75-3788}.

A cavity enables quantum information processes for several
reasons.  First, one may imagine that a photon in a cavity
bounces between its mirrors a large number of times before
leaking out; this number is called the quality factor $Q$.  If
$Q$ is high, one single photon may interact $Q$ times with a
single atom, and if each interaction accomplishes a weak, \QND\
measurement (see Sec.~\ref{sec:resources}), then the measurement strength is
enhanced by $Q$.

But a cavity does more than this.  It also confines the
electromagnetic field into a small volume.  One manifestation
of this is evident in the spontaneous emission of atoms.
Spontaneous emission can be considered as the simultaneous
coupling of an atom to an infinite continuum of modes of the
electromagnetic field.  A cavity makes the coupling to one
particular mode --- the cavity mode --- substantially stronger
than other, free space modes.  This mode is emitted from the
cavity at a rate $\kappa=\omega_0/Q$, where $\omega_0$ is the
resonant frequency of the cavity.  The coupling of the atom to
the cavity mode, $g$, is proportional to $\sqrt{f/V}$.  Here
$f$ is the oscillator strength of the atom, a measure of its
general coupling to electromagnetic fields irrespective of the
cavity, which depends on details such as the size and resonant
frequency of the atom.  The mode-volume of the cavity, $V$, is
a critical parameter to minimize for strong interactions. If
the energy levels of the atom are matched to the cavity photon
energy $\hbar\omega_0$, the rate at which the combined
atom/cavity system emits photons is approximately
$4g^2/\kappa.$   It is possible for this rate to be much larger
than the rate of emission into non-cavity modes, $\gamma$,
leading to a very large resonant \emph{Purcell factor}:
\begin{equation}
\text{Purcell
factor}=\frac{4g^2}{\kappa\gamma}=\frac{3}{4\pi^2}\left(\frac{\lambda}{n}\right)^3\frac{Q}{V},
\end{equation}
where $\lambda/n$ is the wavelength of the emitted photons in
the material of refractive index $n$.  A large Purcell factor
roughly means that when an atom emits a photon, it is very
likely that the emitted photon enters the cavity mode.  This
cavity mode may then be well coupled to a waveguide, which
strongly directs that photon to an engineered destination. This
parameter is critical for a large variety of proposals using
cQED, even those not involving Purcell-enhanced spontaneous
emission of the atom.  The Purcell factor for a resonant
atom/cavity system is also known as the \textit{cooperativity
factor}, and its inverse is known as the critical atom
number\cite{kimble_early}, i.e. the number of atoms in a cavity
needed to have a profound effect on its optical
characteristics.

Large Purcell factors are generally observed in cavities in the
\emph{weak} or \emph{intermediate coupling regime}, also known
as the \emph{bad cavity limit}, in which $\kappa > g$.  This
regime is useful for applications such as single photon
sources, in which the cavity increases the speed, coherence,
and directionality of emitted photons. It is also the
appropriate regime for schemes in which distant qubits are
probabilistically entangled by heralded photon
scattering\cite{early_cirac_entanglement,childress,hybridnjp,wv06}
(as opposed to photon absorption/emission\cite{czkm97}).
However, a variety of schemes are enabled by the \emph{strong
coupling} limit, in which $g \gg \kappa,\gamma$, meaning that
energy oscillates between the atom and the cavity field many
times before it leaks away as cavity loss or emission into
non-cavity modes. The number $\kappa/g$ is known as the
critical photon number, i.e. the number of photons needed in
the cavity to strongly affect the atom\cite{kimble_early}. In
the strong-coupling regime, the atom-cavity system may be
highly nonlinear, introducing remarkable possibilities for
engineering states of the electromagnetic field and its
entanglement with atoms.

Cavity \QED\ impacts every physical proposal discussed in this
review.  Single photon sources enhanced by the Purcell effect
may be critical for quantum computing with photons, and
potentially scalable methods for logic between photonic qubits
may be mediated by a \CQED\ system. Ions and atoms in distant
traps as well as distant self-assembled quantum dots or
nitrogen-vacancy centers may be entangled via \CQED\
techniques. Purcell-enhanced emission may improve the
measurement of electron and nuclear spins, possibly even in the
optically dark system of P:Si.  One of the most striking recent
developments in superconducting qubit systems is the coupling
of these qubits to microwave cavities far into the strong
coupling regime; much farther than any atomic system has been
able to obtain.  This regime is enabled in part by the large
oscillator strengths of superconducting qubits, but more
dramatically by the small cavity mode volumes $V$ available
from the combination of $\mu$m-wide, lithographically
fabricated one-dimensional superconducting waveguides with
centimeter-scale wavelengths\cite{wiringup}. These developments
have enabled researchers to revisit \CQED\ techniques anew and
test the relevant ideas for enabling photon-mediated quantum
computation.

\comment{
The strong coupling regime at optical wavelengths has been
achieved using atoms falling through Fabry-Perot cavities in
vacuum [PRL 80 4157, 90 133602, Science 298 1372], localized in
such cavities [Nature 425 268, 431 1075, 436 87], and very
recently for atoms falling past a microtoroid
cavity\cite{ao-nat-443-671}. These cavities were fabricated on
chip in a large array, holding great promise for scalability.

Localizing an atom in a cavity could conveniently be achieved
by using solid state ``atoms". Strong coupling in the solid
state has been reached only recently with a quantum dot
embedded in a micropillar optical cavity [Nature 432 197] and a
photonic crystal cavity [Nature 432 200]. Such solid state
\CQED\ promises control over ÒatomÓ placement, scalability and
integration, but lacks the isolation from the environment
offered by single atoms. A major breakthrough would be to
combine the these advantages and incorporate Òsolid-state
atomsÓ in a solid state optical cavity. Key challenges will be
fabricating large $Q/V$ cavities compatible with solid state
ÒatomsÓ, alignment of atoms to the mode maximum, and control
over the microscopic environment of the atom-cavity system.

Cavity \QED\ has important applications within many quantum
computing architectures:\emph{eg} coupling superconducting
qubits to single microwave photons on transmission
lines\cite{wa-nat-431-162,sc-nat-445-515}, thereby providing a
means for coupling multiple qubits; implementing optical
nonlinearities at the single photon level for quantum logic
gates \cite{tu-prl-75-4710,mc-sci-303-1992} as well as more
generalized entangling units \cite{de-pra-76-052312}; and
enabling high-efficiency in hybrid spin-photon systems
\cite{li-pra-73-012304}. It seems likely that \CQED\ will have
a central role in future quantum technologies that require
inter-conversion between flying and stationary qubits.}

\section{Single Photons}
\version{Version 1.0, written by Jeremy 1/6/09\\
         Version 2.0, revised by Jeremy 2/17/09\\
         Typesetting issues changed by Thaddeus, 4/2/09\\
         Version 3.0, revised by Jeremy 4/26/09\\
         Version 3.1, mention of optical frequency combs added by Thaddeus 5/25/09}

Realizing a qubit as the polarization state of a photon (horizontal
$|H\rangle\equiv|0\rangle$ and vertical $|V\rangle\equiv|1\rangle$) is appealing since photons are relatively free of the noise that
plagues other quantum systems, and polarization rotations (equivalent to one qubit gates) can be
easily done using \qu{waveplates} made of birefringent
material (whose refractive index is slightly different for the
two polarizations)\cite{ob-sci-318-1567}. Photons also admit encoding of quantum information in other degrees of freedom, including time-bin and path. Of course a
potential drawback is the light-speed propagation of the qubit,
although this is a tremendous advantage in distributing quantum information.

\begin{figure}
\begin{center}
\includegraphics[width=0.45\textwidth]{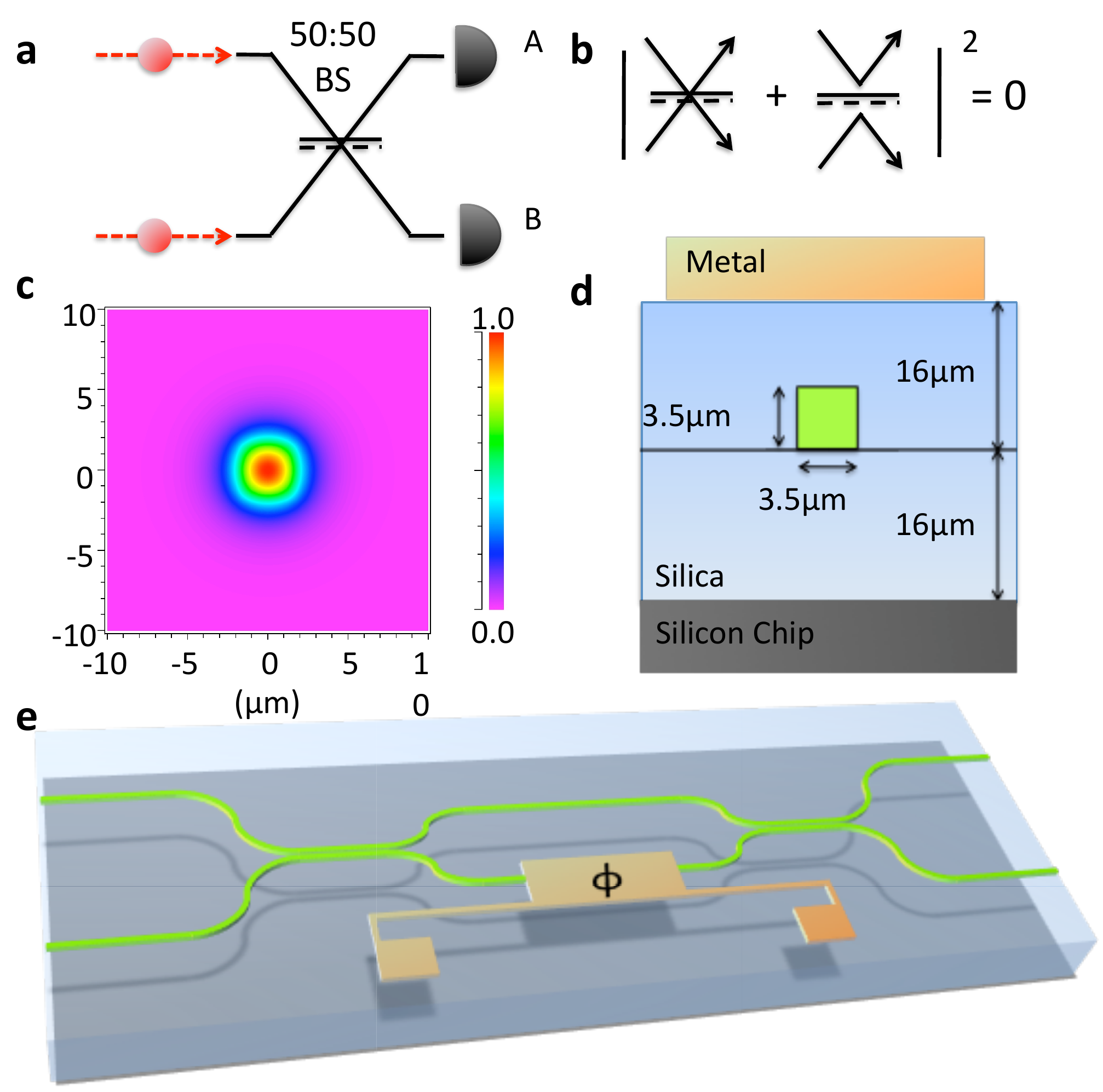}
\caption{Quantum computing with photons. \textbf{a}, Two photons entering a 50:50 beampslitter (50:50 BS) undergo quantum interference. \textbf{b}, The probability amplitudes for the two photons to be transmitted and reflected are indistinguishable and interfere. \textbf{c}, Intensity profile of a photon in a waveguide. \textbf{d}, Silica-on-silicon waveguide structure. \textbf{e}, An interferometer with controlled phase shift for single qubit operations and multi-photon entangled state manipulation.}
\label{photonic}
\end{center}
\end{figure}

A major hurdle for quantum
computing with photons is realising the interactions between
two photons for two-qubit gates. Such interactions require a giant optical
nonlinearity stronger than that available in conventional
nonlinear media, leading to the consideration of
electromagnetically induced transparency
(\abbrev{EIT})\cite{sc-oe-21-1936} and atom-cavity
systems\cite{tu-prl-75-4710}. 
In 2001, a major breakthrough known as
the \abbrev{KLM} scheme showed that scalable quantum computing
is possible using only single-photon sources and detectors, and
linear optical circuits\cite{kn-nat-409-46}. It relied on quantum interference of photons at a beamsplitter (see Fig. \ref{photonic}a,b) to achieve nondeterministic interactions.

Although the \abbrev{KLM} scheme was mathematically shown to be
\qu{in-principle} possible, initially few people believed it was a
`practical' approach, owing to the large resource overhead
arising from the nondeterministic interactions and the
difficulty of controlling photons moving at the speed of light.
This situation has changed over the past five
years\cite{ob-sci-318-1567}: Experimental proof-of-principle
demonstrations of two-\cite{ob-nat-426-264,ob-prl-93-080502,pi-pra-68-032316,ga-prl-93-020504} and three-qubit gates \cite{la-nphys-5-134},
were followed by demonstrations of simple-error-correcting
codes\cite{pi-pra-71-052332,ob-pra-71-060303,lu-pnas-08122008} and simple quantum algorithms\cite{lu-prl-99-250504,la-prl-99-250505}. New theoretical
schemes, which dramatically reduced the considerable resource
overhead\cite{yo-prl-91-037903,ni-prl-93-040503,br-prl-95-010501,ra-prl-95-100501}
by applying the previously abstract ideas of measurement-based
quantum computing\cite{rb01}, were soon followed by
experimental demonstrations\cite{wa-nat-434-169,pr-nat-445-65}.
Today, research efforts are focussed on quantum circuits that can be
fabricated on the chip-scale\cite{po-sci-320-646}, high efficiency single photon detectors \cite{spd} and sources\cite{sps}, and devices
that would enable a deterministic interaction between photons\cite{tu-prl-75-4710}.

The photonic quantum circuits described above were constructed from
large-scale (cm's) optical elements bolted to large optical
tables. While suitable for proof-of-principle demonstrations,
this approach will not lead to miniaturized and
scalable circuits, and is also limited in performance due to
imperfect alignment for quantum interference (Fig. \ref{photonic}a,b).
 Recently it has been
demonstrated that waveguiedes on chip (Fig. \ref{photonic}c,d), which act much like
optical fibres, can be used to implement these
circuits\cite{po-sci-320-646}, and that integrated phase shifters can be used for one-qubit gates and manipulating entangled states on-chip\cite{matthews-2008} (Fig. \ref{photonic}e). Laser direct-write techniques
are also being pursued for three-dimensional circuits\cite{marshall-2008}. Future challenges include developing large-scale circuits with fast switching and integrating them with sources and detectors.

Ideal single photon detectors have high efficiency, high
counting rate, low noise, and can resolve the number of photons
in a single pulse\cite{spd}. Commercial silicon single photon
detectors (Si-\abbrev{APD}s) have have a peak intrinsic
efficiency  of $\sim$70\% and (like photomultipliers) cannot
discriminate between one or more photons. However, work is
being done to increase efficiency and achieve photon number
resolution with
Si-\abbrev{APD}s\cite{ts-ao-46-1009,ka-nphot-2-425}, which
offer room-temperature operation and semiconductor integration.
Semiconductor visible light photon counters (VLPCs) operate at
cryogenic temperatures, have photon number resolution and high
efficiency, but generate a relatively large amount of
noise\cite{ta-apl-74-1063}. Nano-wire superconducting
single-photon detectors absorb a single photon to create a
local resistive \qu{hotspot}, detected as a voltage pulse. The
temperature change, and consequently the voltage change depends
on the absorbed energy.  As a result, the number of photons can
be resolved. Low noise and high efficiency (95\%) have been
achieved for tungsten-based devices
\cite{ir-apl-66-1998,li-oe-16-3032}, although they require
cooling below the critical temperature of 100~mK and are
relatively slow. Superconducting detectors based on
nanostructured NbN are fast (100s MHz), low noise, sensitive
from visible wavelengths to far into the infrared, have
achieved efficiencies of 67\% and photon number
resolution\cite{go-apl-79-705,mi-apl-92-061116,di-nphot-2-302}.

An ideal single photon source is triggered,
frequency-bandwidth-limited, emits into a single
spatio-temporal mode, and has high repetition rate. These
exacting requirements suggest the need for a single quantum
system that emits photons upon transition from an excited to a
ground state. (The excited and grounds states could themselves
be used to encode a qubit, and in fact many of the qubits
described in the following sections have been used to emit
single photons.) Controlling the emission can most conveniently
be achieved by coupling the system to a high-$Q$ optical cavity
(see Sec.~\ref{sec:cqed}); emission of single photons from single atoms has
been demonstrated in this way
\cite{ku-prl-89-067901,mc-sci-303-1992,hi-nphys-3-253}. A
technical difficulty is holding the atom in the optical cavity,
leading to solid state \qu{atom} approaches, such as quantum
dots, and nitrogen vacancies (\NV{s}) in diamond (see Sec.~\ref{sec:diamond})
\cite{sps,sh-nphot-1-215} embedded in semiconductor
microcavities (see Sec.~\ref{sec:cqed}).  A key challenge in these
solid-state sources is to maintain the indistinguishability of
the generated photons\cite{sf02}, which is difficult in
solid-state sources due to spectral jumps and other effects. An
alternative approach is to use the non-linear optical materials
currently used to emit pairs of photons spontaneously:
detection of one photon heralds the generation of the other,
which can in principle be switched into an optical delay or
multiplexed\cite{mi-pra-66-053805}.

While the \abbrev{KLM} and subsequent schemes circumvent the
need for deterministic interactions between photons there are
several schemes for such interactions involving atom-cavity
systems\cite{pe-prl-75-3788,dk04}, which are similar to
approaches to single photon sources (see Sec.~\ref{sec:cqed}). Pioneering
work showed that atom-cavity systems can be used to implement
an optical nonlinearity between photons \cite{tu-prl-75-4710}.
It has been shown that such an atom-cavity system is capable of
implementing arbitrary deterministic interactions
\cite{de-pra-76-052312,st-pra-78-032318}.

The photonic approach to quantum computing remains a leading
one.
(Related approaches based on encoding quantum information in
the continuous phase and amplitude variables of
continuous-wave\cite{br-rmp-77-513} or
mode-locked\cite{me-prl-101-130501} laser beams offer some key
advantages, but these are beyond the scope of this review.)
Achieving scalability will depend on advancements in
waveguides, single-photon sources, and detectors, but whatever
the future holds for photonic quantum computing, it is clear
that photons will continue to play a key role as an information
carrier in quantum technologies.


\section{Trapped Atomic Ions}
\version{Version 2.0, added by Chris, minor edits by Thaddeus 4/4/09\\
         Version 2.1, edited by Chris, 5/28/09}

The best time and frequency standards are based on isolated
atomic systems, owing to the excellent coherence properties of
certain energy levels within atoms\cite{Clocks}.  Likewise,
trapped atoms are among the most reliable type of quantum bit.
Trapped atom qubits can also be measured with nearly $100\%$
efficiency through the use of state-dependent fluorescence
detection\cite{Detect,Acton}.  Current effort with atomic
qubits concentrates on the linking of atoms in a controlled
fashion for the generation of entanglement and the scaling to
larger numbers of qubits.

\begin{figure}
\includegraphics[width=0.45\textwidth]{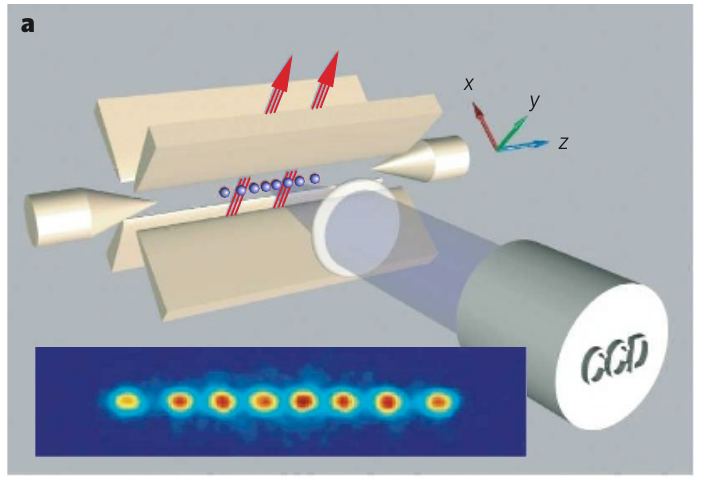}
\caption{Schematic of ion trap apparatus.  Electric potentials are applied to appropriate electrodes
in order to confine a 1-D crystal of individual atomic ions.  Lasers affect coherent spin-dependent
forces to the ions that can entangle their internal qubit levels through their Coulomb-coupled motion.
Resonant lasers can also cause spin-dependent flourescence for the efficient detection of
the trapped ion qubit states.  The inset shows a collection of atomic Ca$^+$ ions fluorescing
(courtesy R. Blatt, University of Innsbruck).}
\label{iontrap}
\end{figure}

Trapped atomic ions are particularly attractive quantum
computer architectures, because the individual charged atoms
can be confined in free space to nanometer precision, and
nearby ions interact strongly through their mutual Coulomb
repulsion\cite{NIST,WinelandBlatt08}. A collection of atomic
ions can be confined with appropriate electric fields from
nearby electrodes, forming a 3-D harmonic confinement
potential, as depicted in Fig.~\ref{iontrap}. When the ions are
laser cooled to near the center of the trap, the balance
between the confinement and the Coulomb repulsion forms a
stationary atomic crystal.  The most typical geometry is a 1-D
linear atomic crystal, where one dimension is made
significantly weaker than the other two\cite{NIST}. In such a
linear trap, the collective motion of the ion chain can be
described accurately by quantized normal modes of harmonic
oscillation, and these modes can couple the individual ions to
form entangled states and quantum gates.

Multiple trapped ion qubits can be entangled through a
laser-induced coupling of the spins mediated by a collective
mode of motion in the trap.  Laser interactions can be used to
simply flip the state of the qubit, or more generally flip the
state of the qubit while simultaneously changing the quantum
state of collective motion.  Such a coupling arises due to
effective frequency modulation of a laser beam in the rest
frame of the oscillating ion and the dipole force from the
laser electric field gradient.  We label the internal qubit
states of ion $i$ as $\up_i$ and $\down_i$, the quantum state
of a Coulomb-coupled mode of collective motion (e.g., the
center-of-mass mode) as $\ket{n}_m$, where $n$ is the harmonic
vibrational index of motion of that particular mode. By driving
ion $i$ on a first order frequency-modulated sideband of the
spin-flip transition, the ion system will undergo Rabi
oscillations between $\down_i\ket{n}_m$ and $\up_i\ket{n \pm
1}_m$, where the plus sign denotes the upper sideband and the
minus demotes the lower sideband\cite{RMP}. We assume that the
sidebands are sufficiently resolved, or equivalently that the
Rabi frequency of the transition is small compared with the
frequency of motion.

The simplest realization of this interaction to form entangling
quantum gates was first proposed\cite{CZ} in 1995 and
demonstrated in the laboratory later that year\cite{CZexp}. The
Cirac-Zoller gate maps a qubit from the the internal levels
within a single trapped ion to the external levels of harmonic
motion, and similarly applies a laser interaction to affect a
second trapped ion qubit conditioned upon the state of motion.
The entangling action of the Cirac-Zoller gate can easily be
seen by considering two successive laser pulses to the two ions
in turn.  We start with the ion pair in the state
$\down_1\down_2\ket{0}_m$ through optical pumping of the qubits
and laser cooling to the ground state of motion.  The first
laser pulse is tuned to drive on the first upper sideband of
the first ion, for a duration that is half of the time required
to completely flip the spin (a $\pi/2-$pulse), and the laser
pulse then drives on the first lower sideband of the second
ion, for a duration set to the time required to completely flip
the spin (a $\pi-$pulse):
\begin{align}
\down_1\down_2\ket{0}_m
    \xrightarrow{\text{pulse 1}}&
        \down_1\down_2\ket{0}_m + \up_1\down_2\ket{1}_m \\
    \xrightarrow{\text{pulse 2}}& \down_1\down_2\ket{0}_m + \up_1\up_2\ket{0}_m \nonumber \\
        &= \left(\down_1\down_2 + \up_1\up_2\right)\ket{0}_m.
\end{align}
These laser interactions entangle the trapped ion qubits, while
the final quantum state of motion is unchanged from its initial
condition.

Extensions to this approach rely on optical spin-dependent
forces that do not require individual optical addressing of the
ions or the preparation of the ions a pure quantum state, and
are thus favored in current experiments\cite{WinelandBlatt08}.
There are also proposals to use radiofrequency magnetic field
gradients \cite{rf_gate} or ultrafast spin-dependent optical
forces \cite{GZC} that do not even require the ions to be
localized to under an optical wavelength (the Lamb-Dicke
limit).

\begin{figure}
\includegraphics[width=0.45\textwidth]{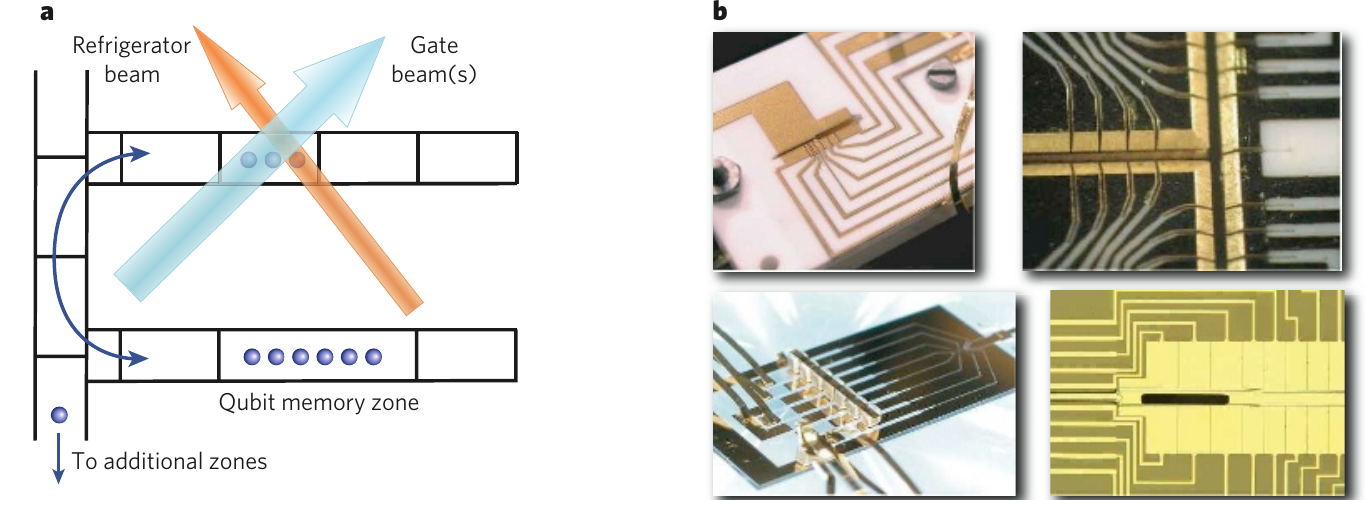}
\includegraphics[width=0.45\textwidth]{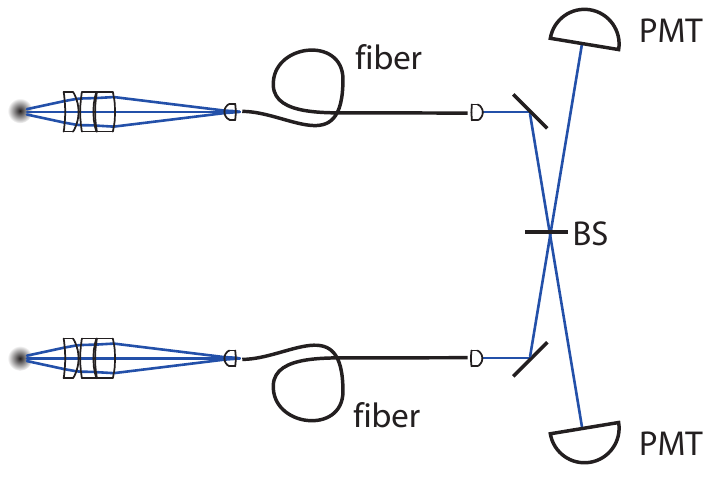}
\caption{Ion trap multiplexing. (a) Entanglement can be propagated to larger collections of trapped ions
by performing quantum gates on small collections of ions (where the motion is under quantum control) and
then physically shuttling the ions to different trapping regions.  (b) This approach may require
more advanced trapping structures that will likely be fabricated on chip structures (courtesy, D. J. Wineland, NIST).
(c) Atoms can be entangled over remote distances through the emission, interference, and detection of photons, depicted with a beamsplitter (BS) and photomultiplier detectors (PMT).}
\label{QCCD}
\end{figure}

The scaling of trapped-ion Coulomb gates becomes difficult when
large numbers of ions participate in the collective motion for
several reasons:  laser-cooling becomes inefficient, the ions
become more susceptible to noisy electric fields and
decoherence of the motional modes\cite{RMP}, and the
densely-packed motional spectrum can potentially degrade
quantum gates through mode crosstalk and
nonlinearities\cite{NIST}.  One promising approach to
circumvent these difficulties is the \qu{Quantum
\abbrev{CCD}}\cite{QCCD}, where individual ions can be shuttled
between various zones of a complex trap structure through the
application of controlled electrical forces from the trap
electrodes, as depicted in Fig.~\ref{QCCD}a.  In this
architecture, entangling gates are operated on only a small
number of ions (perhaps 5--10), where the collective motional
modes can be cold and coherent.  Because the motional state
factors from gate operations, the ions can be moved to
different locations to propagate the entanglement. Auxiliary
ions, perhaps of a different species, can be used as
refrigerators to quench the residual shuttling motion of the
ions through sympathetic laser cooling\cite{WinelandBlatt08}.
There has been great progress in recent years in the
demonstration of multizone ion traps and chip ion traps
(Fig.~\ref{QCCD}b) \cite{Rowe,GaAs,surface,Tee,Cross}.

Another method for scaling ion trap qubits is to couple small
collections of Coulomb-coupled ions through photonic
interactions, as shown in Fig. \ref{QCCD}c. Photonic ion trap
networking offers the significant advantage of having a
communication channel that can easily traverse large distances,
unlike the phonons used in the Coulomb-based quantum gates.
While other matter qubits such as quantum dots and
optically-active impurities can also be coupled in this way,
the use of atoms has the great advantage of reproducibility:
each atom or ion in the network has almost exactly the same
energy spectrum and optical characteristics.  Recently, single
atomic ions have been entangled with the polarization or
frequency of single emitted photons, allowing the entanglement
of ions over macroscopic distances
\cite{Moehring07,Olmschenk09}.  This type of protocol is
similar to probabilistic linear optics quantum computing
schemes discussed above \cite{kn-nat-409-46}, but with the use
of stable qubit memories in the network, this system can be
efficiently scaled to large distance communication through
quantum repeater circuits, and can moreover be scaled to large
numbers of qubits for distributed probabilistic quantum
computing \cite{duanQIC,cluster}.

\section{Neutral Atoms and Optical Lattices}
\version{Version 1.0, added by Chris 5/28/09}

\begin{figure}
\includegraphics[width=0.45\textwidth]{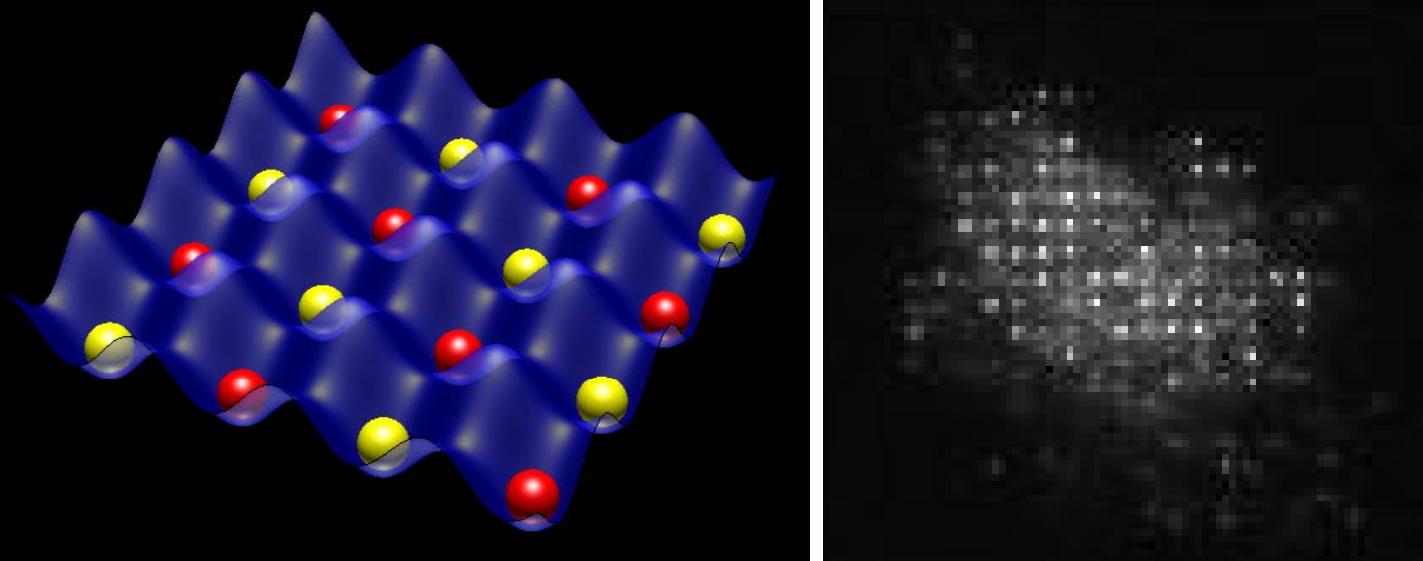}
\caption{(a) Optical Lattice of cold atoms formed by multi-dimensional optical standing wave potentials (courtesy J. V. Porto, NIST). (b) Image of atoms confined in an optical lattice (courtesy D. Weiss, Penn State University).}
\label{lattice}
\end{figure}

A natural host of neutral atoms for quantum information
purposes is the optical lattice - an array of cold atoms
confined in free space by a pattern of crossed laser beams
\cite{LatticeRMP}.  The lasers are typically applied far from
atomic resonance, and the resulting ac Stark shifts in the
atoms results in an effective external trapping potential for
the atoms that is proportional to the squared optical electric
field amplitude.  For appropriate standing wave laser beam
geometries, this can result in a regular pattern of potential
wells in any number of dimensions, with lattice sites spaced by
roughly an optical wavelength (Fig. \ref{lattice}).  Perhaps
the most intriguing aspect of optical lattices is that the
dimensionality, form, depth, and position of optical lattices
can be precisely controlled through the geometry, polarization,
and intensity of the external laser beams defining the lattice.
The central challenges in using optical lattices for quantum
computing are the controlled initialization, interaction, and
measurement of the atomic qubits.  However, there has been much
recent progress on all of these fronts in recent years.

Optical lattices are typically loaded with $10^3$-$10^6$
identical atoms, typically with nonuniform packing of lattice
sites for thermal atoms. However, when a Bose condensate is
loaded in an optical lattice, the competition between intrasite
tunnelling and the on-site interaction between multiple atoms
can result in a Mott-insulator transition where the same number
of atoms (e.g., one) reside in every lattice site
\cite{MI,LatticeRMP}.  Given this external initialization of
the atomic qubits, the initialization and measurement of
internal atomic qubit states in optical lattices can in
principle follow exactly from optical pumping and fluorescence
techniques in ion traps described above.

The interaction between atomic qubits in optical lattices can
be realized in several ways. Optical lattice potentials can
depend upon the internal qubit level (e.g., one state's valley
can be another state's hill), so that atoms in lattices can be
shifted to nearly overlap with their neighbors conditioned upon
their internal qubit state through a simple modulation of the
lattice light polarization or intensity.  Adjacent atoms can
thus be brought together depending on their internal qubit
levels, and through contact interactions, entanglement can be
formed between the atoms.  This approach has been exploited for
the realization of entangling quantum gate operations between
atoms and their neighbors, as depicted in (Fig \ref{lattice}a)
\cite{NIST-Lattice}.  Another approach exploits the observation
that when atoms are promoted to Rydberg states, they possess
very large electric dipole moments.  The Rydberg \qu{dipole
blockade} mechanism prevents more than one atom from being
promoted to a Rydberg state, owing to the induced level shift
of the Rydberg state in nearby atoms \cite{RydbergGate}.  This
effect therefore allows the possibility of controlled
interactions and entanglement.  Recently, the Rydberg blockade
effect was observed in exactly two atoms confined in two
separate optical dipole traps \cite{Wisc,Paris}, and it should
be possible to observe this between atoms in an optical
lattice.

Applying optical lattices to quantum computing involves a
general tradeoff in the atom spacing.  With the natural spacing
of order the wavelength of light, the atoms are close enough
for large interactions, but they are too close to spatially
resolve for individual initialization and addressing.  On the
other hand, larger optical lattice spacings allow the
individual addressing and imaging of the atoms (Fig
\ref{lattice}b), at the expense of much smaller interactions
for the generation of entanglement.  In any case, optical
lattices continue to hold great promise for the generation of
large-scale global entangled states that could be exploited in
alternative quantum computing models, such as cluster-state
quantum computing\cite{rb01}.

\section{Nuclear Magnetic Resonance}
\version{%
  Revision 2.0, written by Ray, added 3/17
\\Revision 2.1, description of benchmarking moved to Box 1.
\\Revision 2.2, modifications by Thaddeus, 4/22
\\Revision 2.3, Ray's references added
}

More than 50 years after its
discovery\cite{purcell:qc1945a,bloch:qc1946a}, research in
nuclear magnetic resonance research is still bringing new
insight on quantum dynamics and control.
In 1996, Cory et al.\cite{cory:qc1997a} as well as Gershenfeld
and Chuang\cite{GC97a} suggested how to use the nuclear spins
in a liquid to build a quantum processor. The idea sprang from
the realization that nuclear magnetic moments are well suited
to bear quantum information for several reasons.  They can be
idealized as two level systems, isolated from their
environment, and controlled with relative ease, taking
advantage of the many years of engineering developed in
\abbrev{MRI} and related technologies.


Immersed in strong magnetic field, nuclear spins can be
identified through their Larmor frequency.  In a molecule,
nuclear Larmor frequencies vary from atom to atom due to
shielding effects from electrons in molecular bonds.
Irradiating the nuclei with resonant radio-frequency (\RF)
pulses allows manipulating them one at a time, inducing generic
one-qubit gates.  Two qubit interactions are implemented using
the indirect coupling mediated through electrons.  In the
liquid state, the rapid tumbling of the molecules effectively
cancels the direct dipolar coupling between nuclei, which is
especially important for eliminating intermolecular
interactions.  Measurement is achieved by observing the induced
current in a coil surrounding the sample of an ensemble of such
qubits.

The other required element is to prepare a fiducial state to
initiate the information processing. It was suggested to turn a
thermal state into a \textit{pseudo-pure} state, i.e. an
ensemble consisting of the desired initial pure state and the
total mixed one.  It was quickly noticed that the proposed
procedure was exponentially inefficient. The problem was
resolved, at least in theory, through the discovery of
algorithmic cooling\cite{schulman:qc1998a,SMW05a}. The use of
highly mixed states also raised questions about the quantumness
of \NMR\cite{braunstein:qc1999a} and the origin of the power of
quantum computers. This spurred research leading to new models
of computation\cite{kl98} and
algorithms\cite{Miquel:2002nx,shor:qc2008a}, suggesting that
there is quantumness despite the use of high-entropy initial
states.

The exquisite control of liquid-state \NMR\ has allowed the
implementation of small algorithms, providing
proof-of-principle of control of quantum processors. This
improvement came not only because of the dramatic development
of the hardware but also the \qu{software}, i.e. using astute
pulse generation, such as composite pulses or shaped pulses to
make them more precise and robust to imperfection.
The long history of pulse techniques from \NMR\ spectroscopy
and \abbrev{mri} have recently been augmented by the new
quantum information focus.  Examples include strongly modulated
pulses\cite{fortunato:qc2002a} and gradient ascent pulse
engineering (\abbrev{GRAPE})\cite{Khaneja:2005im}.


This improved control allowed \NMR\ quantum computation to
manipulate quantum processors of up to a dozen
qubits\cite{marx:qc2000a,Knill:2001sp,vandersypen:qc2001a,Negrevergne:2006uq}.
Important steps towards the implementation of quantum error
correcting protocols have also been made with \NMR. Despite the
loss of polarization in the preparation of the initial
pseudo-pure states, these experiments showed that there was
sufficient control to demonstrate the fundamental workings of
\QEC, but not yet enough for fault tolerance.

Despite its exquisite control, \NMR\ in the liquid state has
its limitations.
The key problem is the scalability limitation arising from the
inefficiency of pseudo-pure-state preparation.
One direction to address this limitation is to move to
solid-state \NMR. A variety of dynamic nuclear polarization
techniques exist in the solid-state, which partially helps
\NMR's principal limitation to scalability. The lack of
molecular motion allows the use of nuclear dipole-dipole
couplings, which may speed up gates by one or two orders of
magnitudes.  A recent example of a step toward solid-state
\NMR\ quantum computation can be found in implementation of
many rounds of heat bath algorithmic
cooling\cite{schulman:qc1998a,SMW05a} using specially made
crystal of crotonic acid.  Different issues of quantum control
arise for this type of technology, and lessons learned from
solid-state \NMR\ experiments may easily be transferred to the
solid state silicon devices discussed in Sec.~\ref{sec:silicon}, and to other
technologies.  Another possibility to extend solid-state \NMR\
systems is to include electrons to assist in nuclear
control\cite{MMS03a,MM06a}. These techniques have possible
application in the diamond-\NV\ system, to be discussed in Sec.~\ref{sec:diamond}.

Despite its limitations, liquid-state \NMR\ has played and
continues to play an important role in the development of
quantum control.  However, the future of \NMR\ lies in the
solid-state, in low temperatures, and in the ability to better
control electrons and their interactions with the nuclei.  In
this way, the lessons learned in \NMR\ quantum computation
research are merging with the solid-state proposals of the
ensuing sections.

    \begin{figure*}[t]
    \begin{center}
    \includegraphics[width=\textwidth]{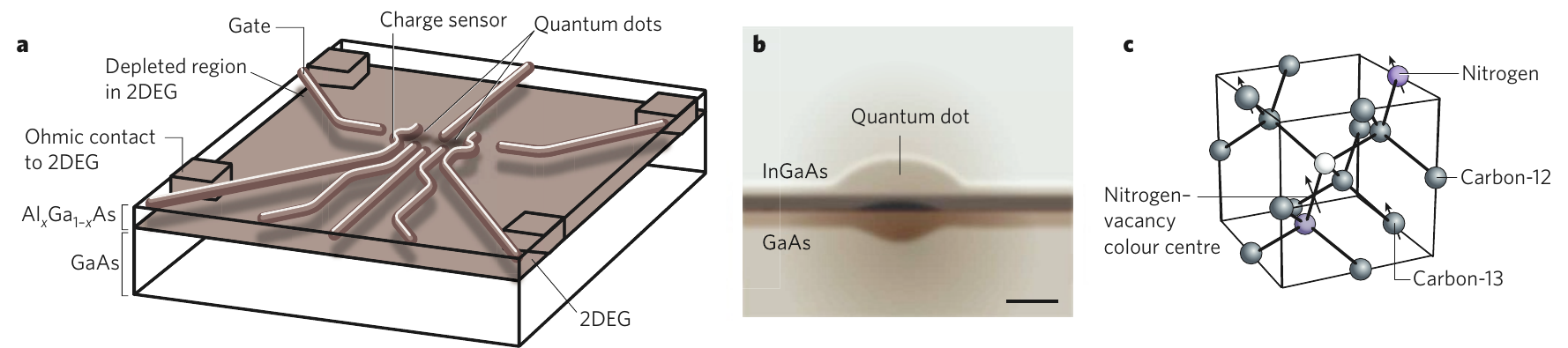}
    \caption{(a) An electrostatically confined quantum dot. (b) A
    self-assembled quantum dot. (c) The atomic structure of a
    nitrogen-vacency center.}
    \label{dotfig}
    \end{center}
    \end{figure*}
\section{Quantum Dots}
\label{sec:dots}
\version{Revision 2.0, written by Thaddeus, added 2/11/09.\\
         Revision 2.1, Thaddeus 3/22/09, shortened CQED discussion based on existence of Box 2.\\
         Revision 2.2, Thaddeus 4/03/09
}

Quantum dots often go by the name \qu{artificial atoms.}  This
terminology highlights their most obvious feature for use in
quantum computing.  They occur when a small nanostructure (in
analogy to a single atomic nucleus) binds one or more electrons
or \qu{holes} (absences of electrons) in a semiconductor.  They
have discrete energy levels that allow coherent control in the
similar ways that trapped ions and neutral atoms are
controlled, and hence their promise for providing useful qubits
is similar.  However, unlike atoms, they do not need to be
cooled and trapped; they are usually born already
integrated into a solid-state host which may be appropriately refrigerated.

Quantum dots come in many varieties, depending on how they are
grown.  In all cases, they confine electrons or holes in a
small region of a semiconductor.  Some quantum dots are
semiconductor nanostructures grown in chemical solution; these
dots are then deposited onto another surface, which may or may
not be another semiconductor.  More common for quantum
computation research are dots grown by molecular beam epitaxy
(\MBE), in which semiconductor crystals are grown layer by
layer, allowing the stacking of different kinds of
semiconductor.  A quantum well is defined by a two-dimensional
plane of a lower bandgap semiconductor (for example, GaAs)
embedded in a larger-bandgap semiconductor (e.g. AlGaAs);
electrons become confined in the lower bandgap layers, which
may be only a few atomic layers thick.  Those electrons might
originate from controlled optical excitation or current
injection.  In some devices they spontaneously \qu{fall} into the
well from a nearby layer of n-type dopants which give up their
electrons for the lower potential of the quantum well.  A
quantum well becomes a quantum dot when an additional
confinement in the remaining two dimensions is added.  Two
important differing classes of quantum dots are
\emph{self-assembled quantum dots}, where a random
semiconductor growth process creates that two-dimensional
confinement, or \emph{electrostatically defined quantum dots},
in which that confinement is defined by electrostatic
potentials created by lithographically fabricated metallic
gates.

One key difference between these two types of quantum dots is
the depth of the atom-like potential they create.
Electrostatically defined quantum dots are typically defined by
small regions in which a two-dimensional electron gas is
depleted.  These dots behave well when the distance electrons
may travel in the two-dimensional electron gas before
scattering is larger than the spatial scale of the structures
defining the dot; these devices therefore require the very low
temperatures ($<$1~K) accessible with dilution refrigerators.
Loading and measuring electrons trapped in these dots is
accomplished by dynamically altering the dot potential by
changing gate voltages.  Self-assembled quantum dots, in
contrast, typically trap electrons with energies much larger
than thermal energies at temperatures several times larger than
a bath of liquid helium ($4$~K).  Their potentials may be
electrically controllable but coherent manipulation is
generally performed using optical rather than electrical
techniques.

One of the earliest proposals for quantum computation in
semiconductors, that of Loss and DiVincenzo,\cite{ld98}
suggested the use of electrostatically defined quantum dots,
whose key advantages over self-assembled dots is that their
location on a semiconductor wafer may be carefully designed.
This proposal envisioned arrays of dots each containing a
single electron, whose two spin states provide qubits.  Quantum
logic would be accomplished by changing voltages on the
electrostatic gates to move electrons closer and further from
each other.  As the electron wave functions begin to overlap,
they form molecular-like orbitals.  These depend on electron
spin due to the Pauli exclusion principle, preventing symmetric
spin-states from occupying the same molecular orbital.  This
combination of Coulomb repulsion with quantum mechanical
Fermi-Dirac statistics is known as the exchange interaction,
and in the Loss and DiVincenzo scheme it is tuned to provide
universal quantum logic.  In their proposal, individual
electron spins could be controlled via microwave transitions
tuned to the spin-splitting in a magnetic field, and spin
measurement could occur via spin-dependent tunneling processes,
reminiscent of technologies in modern magnetic memory.

Since this seminal proposal, substantial progress toward these
goals has been reached.  The spin-dependent tunneling
processes\cite{currentrectification,vandersypen_singlet-triplet}
needed for the measurement of single spins in quantum dots were
demonstrated, and such work has since evolved to employ a
\emph{quantum point contact} (\abbrev{qpc}), which is a
one-dimensional constriction in the potential seen by an
electronic current.  This constriction is sufficiently
sensitive that it may be opened and closed by the charge of a
single trapped electron in a nearby quantum dot.  The
\abbrev{qpc} thereby allows the measurement of a single
electron charge; to measure a spin, the ability of a single
electron to tunnel into or out of a quantum dot must be altered
by its spin state.  This has been done by changing the magnetic
field to alter the energy of a single quantum
dot\cite{vandersypen_singlespin}, and by changing the potential
between two-quantum dots\cite{jptylmhg05,pjtlylmhg05}.  The
control of individual spins in these quantum dots has also been
demonstrated via direct generation of microwave magnetic
fields\cite{koppens06} and by applying microwave electric
fields in conjunction with the spin-orbit
interaction\cite{vandersypen_electric-fields}.  These
techniques have allowed measurement of single spin dephasing
($T_2^*$) and decoherence ($T_2$) times by spin-echo
techniques\cite{vandersypen_spin-echo}.  This single-spin
control turns out to not be necessary for quantum computation;
qubits may be defined by clusters of exchanged-coupled spins,
with effective single-qubit logic controlled by the pairwise
exchange interaction\cite{levy02}. The $T_2$ decoherence of a
qubit defined by an exchange-coupled electron-pair was
measured, also using the spin-echo technique\cite{pjtlylmhg05}.
Voltage control of a two-electron qubit by the exchange
interaction has the particular advantage of being fast; the
single-qubit gates accomplished this way occur in hundreds of
picoseconds, which is faster than a direct microwave transition
for a single spin.

Most of the work described so far has occurred in dots made in
group \abbrev{iii-v} semiconductors.  A critical limitation to
these types of quantum dots is the inevitable presence of
nuclear spins in the semiconductor substrate. Their hyperfine
interactions with the quantum-dot electron spins cause a
variety of interactions. Energy exchange between electron spins
and nuclei is important at low magnetic fields, as observed
experimentally\cite{jptylmhg05,vandersypen_singlet-triplet},
but more critical are dephasing effects.  The random
orientation of nuclear spins at even relatively low temperature
creates an effective inhomogeneous magnetic field, which leads
spins to dephase at a rate of $T_2^*\sim$10~ns.   This static
dephasing may be refocused by spin-echo techniques, and may
also be suppressed by recently discovered effects in which electrically induced electron spin flips pump
nuclear spins to alter the hyperfine
gradient\cite{Reilly_Zamboni_Science}.

But decoherence is still limited by the dynamic spin-diffusion
due to nuclear dipole-dipole interactions.  This process has
been known in the field of electron spin resonance for over 50
years\cite{hahn_spin_diffusion}, but has been revisited by
modern quantum information research\cite{ss03}, where it is
found that in GaAs, nuclear spin diffusion should limit
electron spin decoherence times ($T_2$) of a few $\mu$s, close
to the values observed in Refs.~\onlinecite{pjtlylmhg05} and
\onlinecite{vandersypen_spin-echo}.  Suppressing this
decoherence requires either extraordinary levels of nuclear
polarization, or the dynamic decoupling of nuclear spin noise
by rapid sequences of spin rotations\cite{lv98}.  The latter
approach stems from a long history in magnetic
resonance\cite{CarrPurcell,CPMG}, and recent theoretical
developments in this area suggest a promising future for
extending decoherence times due to nuclear spin-diffusion
noise\cite{udd,magical_udd,universal_udd}.

One way to eliminate nuclear spins is to define similar dots in
nuclear-spin-free group-\abbrev{iv} semiconductors (i.e.
silicon and germanium). Many of the accomplishments
demonstrated in GaAs have recently been duplicated in
SiGe-based\cite{SiGe_dot,sjyc05,sige_dot_demo_2008} or
metal-oxide-semiconductor silicon
(\abbrev{mos})-based\cite{hirayama_silicon_dot,hitachi_silicon_dot}
quantum dots, including single electron charge
sensing\cite{simmons07} and the control of tunnel coupling in
double dots\cite{simmons09}.

Given the experimental progress in the development of
electrostatically defined quantum dots, it is natural to ask
what remains to be done to reach the type of quantum computer
envisioned a decade ago\cite{ld98}. Unfortunately, the
demonstrated interactions between these types of quantum dots
are extremely short-range, and suggest a quantum-computer
architecture with nearest-neighbor interactions only.  When
considering the requirements of fault-tolerant \QEC, this
provides a substantial constraint\cite{szkopek06}.  Although
fault-tolerant operation may be reasonable with a dense,
two-dimensional network of neighbor-coupled
qubits\cite{DiVincenzo-2D,rh07}, such a network may not be
possible due to the space required for the electrical leads
required to define each qubit.  It seems inevitable that a
scalable architecture will require the transport of coherent
quantum information over longer distances.  A number of methods
for accomplishing this in electrostatically defined quantum dot
systems have been proposed, for example using the coherent
shuttling of spins in charge-density waves\cite{faulty-taylor}.
However, it remains experimentally uncertain how far spin
coherence may be reliably transferred on a chip.

Photonic connections between quantum dots may ultimately prove
more reliable, and for this reason optically controlled,
self-assembled quantum dots have also undergone substantial
development.  These quantum dots have a few advantages over
atoms.  Besides the lack of need for motional cooling, their
large size increases their coupling to photons (known as the
\qu{mesoscopic enhancement} of the oscillator strength.)  For
some devices they may be electrically pumped into their excited
states\cite{electrical_pump_qdot}, which may have architectural
advantages in future devices.  One potential use for these
quantum dots in quantum information technologies is as a single
photon source, since after optical or electrical pumping they
efficiently emit one and only
photon\cite{sf02,electrical_pump_qdot} which may be used for
such applications as quantum secret sharing\cite{bb84} or
photonic quantum computers as discussed above.

The earliest proposals for the use of optically controlled
quantum dots for quantum computing\cite{iabdlss99,sim99,skh99}
stressed the importance of optical microcavities for allowing
photons to mediate quantum logic between the dots (see Sec.~\ref{sec:cqed}).
Many schemes have been devised, often based on early proposals
for establishing entanglement between atoms\cite{czkm97}.
Recently, it has been realized that the photonic wiring of
quantum-dot-based quantum computers should be possible with
experimentally realistic cavities\cite{yls05,wv06,hybridnjp}.
The strong-coupling regime is particularly challenging in
solid-state settings where surface effects on lithographically
defined microstructures degrade the cavity $Q$.  Most schemes
for optically connecting quantum dots via microcavities and
waveguides only require a high Purcell factor or cooperativity
parameter.  Although solid-state microcavities may have smaller
$Q$, $Q/V$ may be very large due to very small mode volumes, on
the scale of a cube of the optical wavelength. For this reason,
there has been substantial development of systems incorporating
a quantum dot in a microcavity.  There are a variety of
microcavity designs; those that have demonstrated
strong-coupling operation include distributed Bragg reflector
micropillars\cite{Forchel04a},
microdisks/microrings\cite{Gerard05a}, and defects in photonic
bandgap crystals\cite{Gibbs04a,Imamoglu07a}.

The control and measurement of self-assembled quantum dots has
also made recent progress.  This research is hindered by the
random nature of these quantum dots; unlike atoms, their
optical characteristics vary from dot to dot, so many
experiments that work for one device may fail for another.
Nonetheless, rapid optical initialization of spin-qubits in
quantum dots has been demonstrated for both electrons and
holes\cite{Imamoglu_spin-state-prep,hole-pumping,pressnature}.
Optical quantum non-demolition measurements have been
demonstrated\cite{awschalom_dps,imamoglu_dps}, and single-spin
control via ultrafast pulses has been
developed\cite{bayerT2,awschsingle,pressnature}.  A remarkable
feature of this optical control is that these qubits may be
controlled very quickly, on the order of picoseconds,
potentially enabling extremely fast quantum computers.  Initial
demonstrations of quantum logic between single quantum dots in
microcavities and single photons have also
begun\cite{vuckovic_phaseshift}.

Although single qubit preparation, control, and measurement in
single, self-assembled quantum dots are now well established,
substantial challenges remain in scaling to larger systems.
First, the many schemes for establishing entanglement between
quantum dots are either probabilistic or insufficiently robust
to photon loss, which is typically a large problem in realistic
chip-based devices. Second, self-assembly leads to dots that
are randomly placed spatially and spectrally.  Emerging
fabrication techniques for deterministic placement of
dots\cite{forchel_alignment} and dot tuning
techniques\cite{forchel_tuning,vuckovic_phaseshift} may remedy
this problem in the future. Finally, these quantum dots suffer
the same nuclear-spin-induced decoherence issues faced by the
electrostatically defined quantum dots, and will likely require
similar dynamical decoupling methods.  Another approach under
consideration is the use of a hole spin rather than an electron
spin, since in GaAs holes have spatial wavefunctions with
substantially smaller overlap with the nuclear spins, weakening
the effects of this interaction, and potentially extending
decoherence to the lifetime limit\cite{loss-hole}. In bulk
semiconductors, holes typically have much shorter relaxation
times due to stronger spin-orbit relaxation; their utility for
quantum-dot qubits remains to be seen.  Initial results in the
initialization\cite{hole-pumping} and the measurement of long
$T_2^*$ values\cite{hole-CPT} show remarkable promise.  These
and other results suggest a long future for improving the
viability and scalability of optically controlled quantum dots
for large-scale quantum computation.

\section{Impurities in Silicon}
\label{sec:silicon}
\version{%
  Version 0.5, written by Jeremy, added 1/6/09.
\\Version 1.0, revision by Thaddeus 3/24/09
\\Version 1.1, references added by Thaddeus 4/2/09}

In 1998, at the same time as the first demonstrations of
quantum computing in \NMR\ systems were being realised and
close to the appearance of the Loss and DiVincenzo\cite{ld98}
scheme, Bruce Kane developed a proposal to marry \NMR\ quantum
computation with a silicon-based system\cite{ka-nat-393-133}.
The Kane proposal was highly influential, primarily since it
seems to be highly consistent with extant silicon-based
microelectronic technologies.  This proposal embeds quantum
information in the state of nuclear-spin ($I=1/2$) qubits.
However, unlike in liquid-state \NMR\ they are \textit{single}
nuclear spins of individual phosphorus $^{31}$P nuclei embedded
in isotopically pure silicon-28 ($^{28}$Si), which has a
nuclear spin $I=0$. Phosphorus is a standard \textit{donor} in
silicon, donating one electron to attain the same electronic
configuration as silicon. At low temperatures this donor
electron is bound to the phosphorus nucleus.  These donor
electrons are critical to the operation of the quantum
computer: they mediate a nuclear spin interaction, allow qubits
to be addressed individually and are integral to measuring the
spin state of the qubits.

Coincidental to quantum computing, isotopically purified
silicon started to become available\cite{tc_rmp_05}, and bulk
samples of this silicon have now shown remarkable properties
supporting the Kane proposal.  The electron spins in $^{28}$Si
show encouragingly long $T_2$ times, exceeding
60~ms, as demonstrated by electron-spin resonance\cite{tlar03}.
This coherence has recently been extended to a few seconds by
swapping the electron coherence with the $^{31}$P nuclear
spin\cite{mtbslashal08}; the potential for much longer nuclear
spin decoherence times of minutes or longer has further been
seen in \NMR\ dynamic decoupling experiments\cite{lmyai05} on
$^{29}$Si in $^{28}$Si. Another remarkable property of
isotopically purified silicon is that the optical transitions
related to the $^{31}$P donor become remarkably sharp in
comparison to isotopically natural silicon\cite{thewalt}.
Unlike in any other semiconductor to date, the optical
transitions are sharp enough to resolve the hyperfine splitting
due to the $^{31}$P nuclear spin in the optical
spectra\cite{thewalt}.  This has enabled rapid (less than 1
second) electron and nuclear spin polarization by optical
pumping\cite{thewalt_pol}, orders of magnitude faster than the
polarization obtained in 50 years of research into silicon spin
polarization\cite{Feher56polarization,mccamey_pol}.  Rapid
polarization is critical for the success of proposals such as
Kane's, since $T_1$ times (see Sec.~\ref{sec:decoherence}) for electron and nuclear
spins in silicon are notoriously long at low temperature and
qubits must be constantly initialized for \QEC.

Quantum logic in the Kane proposal is similar to the proposal
for quantum dots of Loss and Divincenzo\cite{ld98} discussed in
the previous section.  The wave functions of the electrons
bound to phosphorus impurities are controlled by
nanometer-scale metallic gates, and the resultant
exchange-split energy levels in turn affect the energy levels
of the nuclear spins, due to the strong Fermi contact hyperfine
coupling between the electron spins and the nuclei.  This
effect, in addition to magnetic resonance techniques using
radio-frequency (\abbrev{RF}) magnetic fields, allows universal
control of single spins and nearest-neighbor two-qubit quantum
gates.

Additional ideas for quantum computing in silicon include the
'spin resonance transistor,' in which the varying gyromagnetic
ratio of spins in different semiconductors allow the electrical
control of donor-bound electronic spins in Si/Ge alloys without
the need for \abbrev{RF} fields\cite{VYWJBRMD2000}.
Further departures\cite{lgyyai02} include eliminating the
electronics altogether and computing with arrays of spin-1/2
$^{29}$Si in $^{28}$Si, or using dipolar couplings between
donor-bound electron spins\cite{sdd04}.  Much recent work has
focused on silicon-based quantum dots, as discussed in the
previous section.

\begin{figure}[t]
\begin{center}
\includegraphics[width=0.4\textwidth]{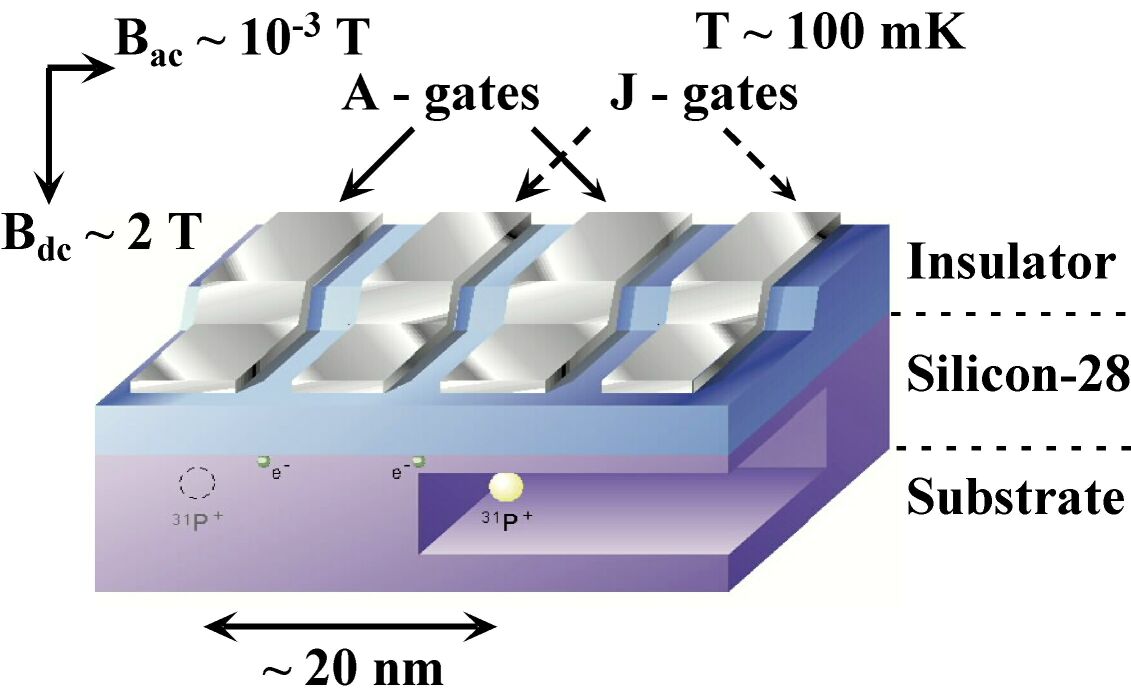}
\caption{Schematic of the original Kane architecture}
\label{hom}
\end{center}
\end{figure}

\comment{
At low temperatures the electrons can be spin polarised by a
modest external magnetic field when the Zeeman energy exceeds
the thermal energy: 2$\mu_BB\gg k_BT$. This gives
$n_\uparrow/n_\downarrow<10^{-6}$ for $T\leqslant$100 mK and
$B\geqslant$2 T \cite{ka-nat-393-133}. The nuclear spins
themselves are not polarized by this field but can be polarised
through interactions with the polarised electrons, as discussed
below.

The donor nuclear spin-electron system with $B\| z$ can be
described by the Hamiltonian:
\begin{equation}\label{Hen}
  H_{en}=\mu_BB\sigma_z^e-g_n\mu_nB\sigma_z^n+A\sigma^e\sigma^n
\end{equation}
where $\sigma$ are the Pauli spin matrices, $\mu_n$ is the
nuclear magneton, $g_n$ is the nuclear $g$-factor and $A$ is
the contact hyperfine interaction given by
\begin{equation}\label{hyperfineA}
  A=\frac{8}{3}\pi\mu_Bg_n\mu_n|\Psi(0)|^2
\end{equation}
where $\Psi(0)$ is the electron wavefunction at the nucleus.
For an electron in its ground state the separation of nuclear
levels is
\begin{equation}\label{freq}
  h\nu_A=2g_n\mu_nB+2A+\frac{2A^2}{\mu_BB}
\end{equation}

It is clear from Eqs. \ref{hyperfineA} and \ref{freq} that the
hyperfine interaction and hence the resonance frequency between
levels of the nucleus can be tuned by controlling the electron
wavefunction overlap with the nucleus. This is achieved by
application of a positive bias to a surface \textit{A}-gate
which pulls the electron wavefunction away from the nucleus
towards the barrier layer (inset to Fig. \ref{kane2}) The
nuclear resonance frequency has been calculated by Kane for
applied A-gate voltage and the results are shown in Fig.
\ref{kane2}. Application of an appropriate $A$-gate voltage can
be used to selectively bring the nuclear spin system into
resonance with a global harmonic magnetic field $B_{ac}$. In
this way any arbitrary single qubit rotation can be performed.

An exchange interaction between nuclear spins, mediated by the
electron spins, can be used for a \CNOT\ gate if the donors are
sufficiently close together. The Hamiltonian for the two donor
electron-nuclear spin system is \cite{ka-nat-393-133}:
\begin{equation}\label{}
  H=H(B)+A_1\sigma^{1n}\sigma^{1e}+A_2\sigma^{2n}\sigma^{2e}+J\sigma^{1e}\sigma^{2e}
\end{equation}
Where the $A$'s are given in Eq. \ref{hyperfineA} and $H(B)$
includes the interactions of the spins with the applied
magnetic field. The exchange energy 4$J$ depends on the
wavefunction overlap:
\begin{equation}\label{}
  4J(r)\cong1.6\frac{e^2}{\varepsilon
  a_B}\left(\frac{r}{a_B}\right)^{\frac{5}{2}}exp\left(\frac{-2r}{a_B}\right)
\end{equation}
where $r$ is the donor separation, $\varepsilon$ the dielectric
constant of silicon and $a_B$ the Bohr radius. The exchange
frequency has been calculated as a function of donor separation
in Ref. \cite{ka-nat-393-133} and is shown in Fig. \ref{kane3}.
This exchange overlap can also be varied by an electrostatic
potential applied to the $J$-gate.

In this design, the nuclear spin states are measured by first
adiabatically converting them to states with different electron
spin orientation. Coupling of this electron spin to a second
electron spin, on a second donor site, with known spin
orientation can be used to measure the unknown spin. This can
be done by electrostatically encouraging the formation of a two
electron state on the second donor (a $D^-$ state) with a
differential bias applied to the respective $A$-gates. The
$D^-$ state is always a spin singlet and will therefore only
form if the two electron spins are of opposite orientations.
The nuclear spin state can then be inferred by the formation or
otherwise of the $D^-$ state. The single spin measurement is
effectively reduced to a single charge measurement. This process
can also be used to initialize the computer by measuring spin
states and flipping them if necessary.
}

The novel quantum logic ideas of silicon quantum computing
proposals have not yet been demonstrated, since single-spin
measurement in this system must push existing nanotechnology
techniques.  Unfortunately, the single-spin measurement
techniques described in the other sections of this review, such
as those for electrically gated quantum dots (Sec.~\ref{sec:dots}),
cannot be easily applied to silicon.   Optical detection of
single spins, as established for self-assembled quantum dots (Sec.~\ref{sec:dots})
and diamond-\NV\ centers (Sec.~\ref{sec:diamond}), is hindered by silicon's indirect
bandgap, requiring heroic improvement by cavity
\QED\cite{flsy04} (Sec.~\ref{sec:cqed}).  Kane's solution to the problem
of measurement begins by coherently transferring the state of a
single nuclear spin to the donor electron, and then
transferring that electron spin to charge by comparing it to
the spin of a nearby donor, again relying on the Pauli
exclusion principle.  Then, single electron charges near a
Si/SiO$_2$ interface must be sensed.

For single-charge sensing in silicon, there is substantial
prior art in the development of silicon-based single electron
transistors (\abbrev{SET}s) operating as extremely sensitive
charge amplifiers \cite{ka-prb-61-2961}.  Charging of
silicon-based quantum dots has been detected by \abbrev{SET}s
operating at \abbrev{RF} frequencies, making critical use of
multiple devices for noise cancellation\cite{brbhdc03}.
Recently, silicon-based transistors have aided the detection of the ion-implantation of single dopants\cite{bwrps07}, a technique which adds to \abbrev{STM} techniques\cite{obrien01} for placing phosphorus impurities in prescribed atomic locations.  Single spin detection has not yet been accomplished, but innovations in electrically detected magnetic resonance (\abbrev{EDMR}) have resolved small ensembles of phosphorous impurity spins\cite{si_edmr_natphys}, and the spin states of single impurities in the oxides of silicon-based field-effect transistors have been successfully detected\cite{jiang_single}.  Some combination of these techniques are likely to achieve single spin detection in the near future.

What existing measurements with \abbrev{SET}s reveal\cite{obrien01}, as well as some \NMR\ data\cite{lmyai05}, is that measurement and decoherence in this system are limited by $1/f$ noise, a familiar noise source in classical silicon-based electronics due to random charge states at silicon/insulator interfaces.  This noise source is no surprise, as indicated by Kane\cite{ka-nat-393-133}.   Reduction of this noise source to the small levels required for fault tolerant quantum computing requires the development of clean, high-quality silicon/insulator/metal interfaces.  This challenge is expected to be surmountable due to silicon's primary advantage: the massive  infrastructure in high-quality silicon microprocessing that already exists for large-scale classical computing.  Despite the challenges in measurement and nanofabrication, silicon-based quantum computers maintain substantial hope of \qu{taking off} due to their ability to leverage existing resources for very large scale integration once the fundamental difficulties are solved.

\section{Impurities in Diamond}
\label{sec:diamond}
\version{Version 1.0, added by Fedor 3/30/09\\
         Version 1.1, slightly revised by Thaddeus 3/30/09\\
         Version 1.2, revised by Thaddeus 4/23/09, 6/8/09}

Diamond is not only the most valuable gemstone, but also an
important material for semiconductor technology. It holds
promise to replace silicon owing to unprecedented thermal
conductivity, high charge carrier mobility, hardness, and
chemical inertness.   Dopants in diamond can be used as a
platform for quantum information processing devices, like the
phosphorus impurities in silicon discussed in the previous
section.

Diamond hosts more than 500 documented optically active
impurities, known as colour centres, since they are responsible
for coloration in crystals.  Nitrogen, being the most abundant
impurity in diamond, forms about ten optically active defects
including the nitrogen-vacancy (\NV) centre. The structure of
the \NV\ centre (shown in Fig.~\ref{dotfig}c) consists of a
substitutional nitrogen at the lattice site neighboring a
missing carbon atom.  It is established experimentally that
these \NV\ centres can exist in two charge states as neutral
and negatively charged.  Several unique properties make the
\NV\ centres particularly suitable for applications related to
quantum information processing. First, the \NV\ center exhibits
strong optical absorption and high fluorescence yield
that allows the detection of a single defect using confocal
fluorescence microscopy\cite{gruberscience} (and recently
developed nonlinear microscopy techniques allow far field
addressing of defects with a resolution of about
5.6\nm\cite{STED_diamond}).  Second, it is extraordinarily
photostable, meaning that it does not show any photoinduced
bleaching upon strong illumination. Third, the paramagnetic
ground state of a charged \NV\ defect can be used as a
qubit\cite{wrachtrup_proposal}. Finally, the fluorescence
intensity of a \NV\ defect is spin-dependent, which allows the
readout of its spin state via counting the number of scattered
photons\cite{jgpdgw04}.

The remarkable properties of the \NV\ centre have already found
application as a single photon source for quantum
cryptography\cite{kmzw00}, including the first commercial
single photon source device available on the market.
Spin-based quantum information processing can also profit from
the outstanding properties of the diamond lattice.

The negatively charged state of the \NV\ centre is formed by
four electrons associated with dangling bonds of the vacancy,
one electron originating from nitrogen, and an additional
electron from an external donor.  Two out of these six
electrons are unpaired forming a triplet spin system. Spin-spin
interactions split the energy levels with magnetic quantum
numbers $m_s=0$ and $m_s=1$ by about 2.88\GHz. The degeneracy
of $m_s=\pm 1$ states, arising from C$_\text{3v}$ symmetry, can
be lifted further by applying an external magnetic field. Under
optical illumination, spin-selective relaxations lead to an
efficient optical pumping of the system into the $m_s=0$ state,
allowing fast (250\ns) initialization of the spin
qubit\cite{hsm06}. The spin state of a \NV\ centre can be
manipulated by applying resonant microwave
fields\cite{jelezko_04a}. Hence all the necessary ingredients
to prepare, manipulate and readout single-spin qubits are
readily available in diamond.
The first demonstration of quantum process tomography in solid
state was realized on a single diamond spin shortly after the
discovery of spin manipulation techniques in this
system\cite{diamond_tomography}.

In contrast to GaAs quantum dots,
spins in \NV\ centres show long decoherence times, even at room
temperature.  The observed decoherence times depend on the
growth method of the diamond lattice.
In low-purity technical grade synthetic material (type 1b
diamond), single substitutional nitrogen atoms cause major
effects on the electronic spin properties of \NV\ centres.
Flip-flop processes from the electron spin bath create
fluctuating magnetic fields at the location of the \NV\ centre
limiting the coherence time to a few microseconds. It was shown
that by applying an external magnetic field, these spin
fluctuations can be suppressed substantially\cite{hdfga08}.
Furthermore, the electron spin bath can be polarized in high
magnetic fields leading to complete freezing of nitrogen spin
dynamics\cite{thvsa08}. Another way to prolong coherence times
comes from the possibility to grow ultrapure diamond. Recently,
it was shown that a chemical vapor deposition process allows
reducing the impurity concentration down to about 0.1 parts per
billion. In such materials, the nuclear spin bath formed by
$^{13}$C nuclei (natural abundance of about 1.1 percent)
governs the dynamics of electron spin of \NV\
centres\cite{gaebel2006}.  The decoherence of electron spins
can be remarkably long if these nuclei are removed.  By growing
isotopically enriched $^{12}$C diamond it is possible to
increase $T_2$ to 2\ms\ for 99.7\% pure
material\cite{diamond_ultralongT2}.

In lattices that do contain $^{13}$C nuclei, it is found that
those nuclear spins located close to the \NV\ centre are
excluded from the spin dynamics owing to an energetic detuning
from the dipolar interaction with the electron spin. These
nuclear spins, located in the \qu{frozen core} extending to
about 4 nanometers from the electron spin, can be initialized
and controlled by the \NV\ centre.  They can themselves be used
as a quantum memory, which may be particularly useful in
quantum repeaters\cite{childress_science}.  For example, the
state of the electron spin can be mapped onto the nuclear spin
state (which phase memory can be as long as seconds) and
retrieved with very high fidelity\cite{dutt_science}.
Three-spin entanglement was also demonstrated for two nuclei
coupled to the electron spin\cite{diamond_entanglement}.


Intrinsic coupling of stationary qubits (spins) to flying
qubits (photons), manifested for example in the effect of
electromagnetically induced transparency
(\abbrev{EIT})\cite{diamond_CPT}, allows coupling between
distant \NV\ centres.  This capability enables quantum
computation schemes based on probabilistic entanglement between
distant qubits\cite{childress}, as discussed above in the
context of trapped ions.  Optical transitions of \NV\ centres
may be sufficiently ``atom-like" in that they are not affected
by dynamic inhomogeneity (i.e. they have a transform-limited
linewidth), potentially enabling interference from two distant
defects. Static inhomogeneity caused by strain present in the
crystal lattice (which is on the order of 30\GHz\ for high
quality synthetic crystals) can be compensated by applying an
external electric field (Stark effect)\cite{diamond_stark}.

Deterministic schemes for creating entanglement between distant
spin qubits via a photonic channel require coupling of optical
transitions to a high-$Q$ cavity (see Sec.~\ref{sec:cqed}).  The first
experimental demonstration of such coupling was reported for
whispering gallery modes of silica
microspheres\cite{diamond_microsphere}.  More recently,
monolithic diamond photonic structures were designed and
fabricated, including waveguides and photonic crystal
cavities\cite{diamond_pc_2006,diamond_micromachine}. When
incorporated into photonic structures, diamond defects can
provide the platform for an integrated quantum information
toolbox, including single photon sources and quantum memory
elements.

Many initial benchmark demonstration experiments on coherent
control of a diamond quantum register were carried out on
naturally formed \NV\ centres. However, for many applications,
in particular those related to coupling of \NV\ centres to
optical cavities, it is necessary to control the position of
\NV\ centres. Although creation of \NV\ centres in
nitrogen-rich diamond by electron irradiation is an established
technique, its poor positioning accuracy is not suitable for
quantum information devices. Recently, implantation techniques
relying on atomic and molecular implantation of nitrogen in
ultrapure diamond using focused ion beams were
reported\cite{gaebel2006}. Although generation of \NV\ defects
remains probabilistic owing to fluctuation of the ion number in
the beam, novel approaches involving cold ion traps as a source
are also proposed.  Note that use of single cold ions not only
eliminates statistical fluctuation of the number of implanted
ions, but also allow {\AA}ngstrom-level accuracy of positioning
them into crystal\cite{diamond_ion_concept,ion_targetting}.

While most of the quantum information processing work was
performed on \NV\ centres, new emerging systems based on
nickel- and silicon-related defects were also reported
recently\cite{wu07,nickelnitrogen05,SiV06}. Optical properties
of nickel-related centres outperform \NV\ centres owing to
their narrow-band, near-infrared emission at room temperature
which is important for free-space and fiber-based quantum
communication. The silicon-vacancy defect is particularly
interesting because it is known to have paramagnetic ground
electron state similar to \NV\ defects. Therefore it is likely
that other defect centers in addition to \NV\ centres have
strong potential for use in quantum information technology.


\section{Superconducting Qubits}
\version{Version 2.0, written by Thaddeus 2/3/09\\
        Revision 2.1.  Submitted by Yasu 3/4.\\
        Revision 2.2, minor changes by Thaddeus, old references added 3/31/09\\
        Revision 2.3, minor changes by Yasu, new references added 4/6/09\\
        Revision 2.5, minor changes by Thaddeus 4/20/09\\
        Revision 2.6, changes by Yasu, some references removed 5/12/09\\
        Revision 2.7, changes by Yasu to section about decoherence}

%
%
%
%


If you tried to make a quantum computer using classical
electronics, you would find that the resistance of normal
metals would constantly leak the quantum information into heat,
causing rapid decoherence. This problem may be alleviated using
zero-resistance superconducting circuits.


The basic physics behind superconducting qubits is most easily
explained by analogy to the simpler quantum mechanical system
of a single particle in a potential. To begin, an ordinary
LC-resonator circuit provides a quantum harmonic oscillator.
The magnetic flux across the inductor $\Phi$ and the charge on
the capacitor plate $Q$ have the commutator $[\Phi,Q]=i\hbar$,
and therefore $\Phi$ and $Q$ are respectively analogous to the
position and momentum of a single quantum particle. The
dynamics are determined by the \qu{potential} energy
$\Phi^2/2L$ and the \qu{kinetic} energy $Q^2/2C$, which results
in the well-known equidistant level quantization of the
harmonic oscillator. However, this level structure does not
allow universal quantum control.
Anharmonicity is needed, which is available from the key
component in superconducting qubits: the Josephson junction.
A Josephson junction is a thin insulating layer separating sections
of the superconductor, in which quantum tunneling
of Cooper pairs may still occur. 
The quantization of the tunneling charge across the junction brings
a cosine term in the potential energy.
Thus, the total potential in the parallel circuit shown in Fig.~7a is
\begin{equation}
    U(\Phi) = E_J \left[ 1 - \cos \left( 2\pi \frac{\Phi_{\rm ex}-\Phi}{\Phi_0} \right) \right] + \frac{\Phi^2}{2L} ,
\end{equation}
in terms of the flux quantum $\Phi_0 = h/2e$ and the Josephson
energy $\tsc{E}{J}$, which is proportional to the junction
critical current.
Two of the quantized levels in the anharmonic potential $U(\Phi)$ give rise
to a qubit.


There are three basic types of superconducting qubits, {\it
charge}, {\it flux}, and {\it phase}, which are conveniently
classified by the bias flux $\Phi_{\rm ex}$. The ratio
$E_J/E_C$ is also crucial, where $E_C = e^2/2C$ is  the single
electron charging energy characterizing the charging effect,
i.e. the kinetic term.

The {\it charge} qubit
omits the inductance. There is no closed superconducting loop,
and the potential is simply a cosine one with a minimum at zero
phase. It is sometimes called a \textit{Cooper-pair box}, as it
relies ultimately on the quantization of charge into individual
Cooper pairs, which becomes a dominant effect when a
sufficiently small \qu{box} electrode is defined by a Josephson
junction. Qubits of this type were first proposed
\cite{buttiker87,ssh97}                         
and developed
\cite{
      saclaydemo98,
      npt99}            
in the regime of $E_J/E_C \ll 1$, and later extended to the
other limit and named \textit{quantronium}\cite{vacjpued02} and
\textit{transmon}\cite{Houck08}. The nature of the wave
functions and their sensitivity to charge fluctuations depend
critically on the choice of $E_J/E_C$.

In the {\it flux} qubit
\cite{Leggett80,
      moltvl99,%
      cnhm03},%
, also known as a \textit{persistent-current qubit}, $\Phi_{\rm
ex} \simeq \Phi_0/2$ is chosen to give a double-well potential.
The two minima correspond to persistent current going in one
direction along the loop or the other. Often, the inductance is
substituted by an array of Josephson junctions. The kinetic
energy term is kept small, $E_J/E_C \gg 1$.

In the {\it phase} qubit%
    \cite{
       mnau02}%
, the potential is biased at a different point, for example
$\Phi_{\rm ex} \simeq \Phi_0/4 $, and again $E_J/E_C \gg 1$.
Unlike the flux qubit, the phase qubit uses the two-lowest
energy states in a single metastable potential well which is
anharmonic.

\newcommand{\thiscaption}{
  (a) Minimal circuit model of superconducting qubits. Josephson junction is denoted by \textsf{X}.  The capacitance \textsf{C} includes a contribution from the junction itself.
  (b)-(d) Potential energy $U(\Phi)$ (red) and qubit energy levels (black) for
  (b) charge, (c) flux, and (d) phase qubit, respectively. The potential for charge qubit is under a periodic boundary condition.
  (e)-(h) Micrographs of superconducting qubits. The circuits are made of Al films. The Josephson junctions consist of Al$_2$O$_3$ tunnel barrier between two layers of Al.
  (e) Charge qubit, or a Cooper pair box.
  (f) Transmon, a derivative of charge qubit with large $E_J/E_C$.
      The Josephson junction in the middle is not visible in this scale.
      The large interdigitated structure is a shunt capacitor. 
  (g) Flux qubit. Two of the three junctions in the series provide inductance. 
  (h) Phase qubit. 
}
\newcommand{\thisfigure}{figs/superconducting_qubits.jpg}

\begin{figure*}
\ifthenelse{\boolean{ispreprint}}
{
    \includegraphics[width=0.66\textwidth]{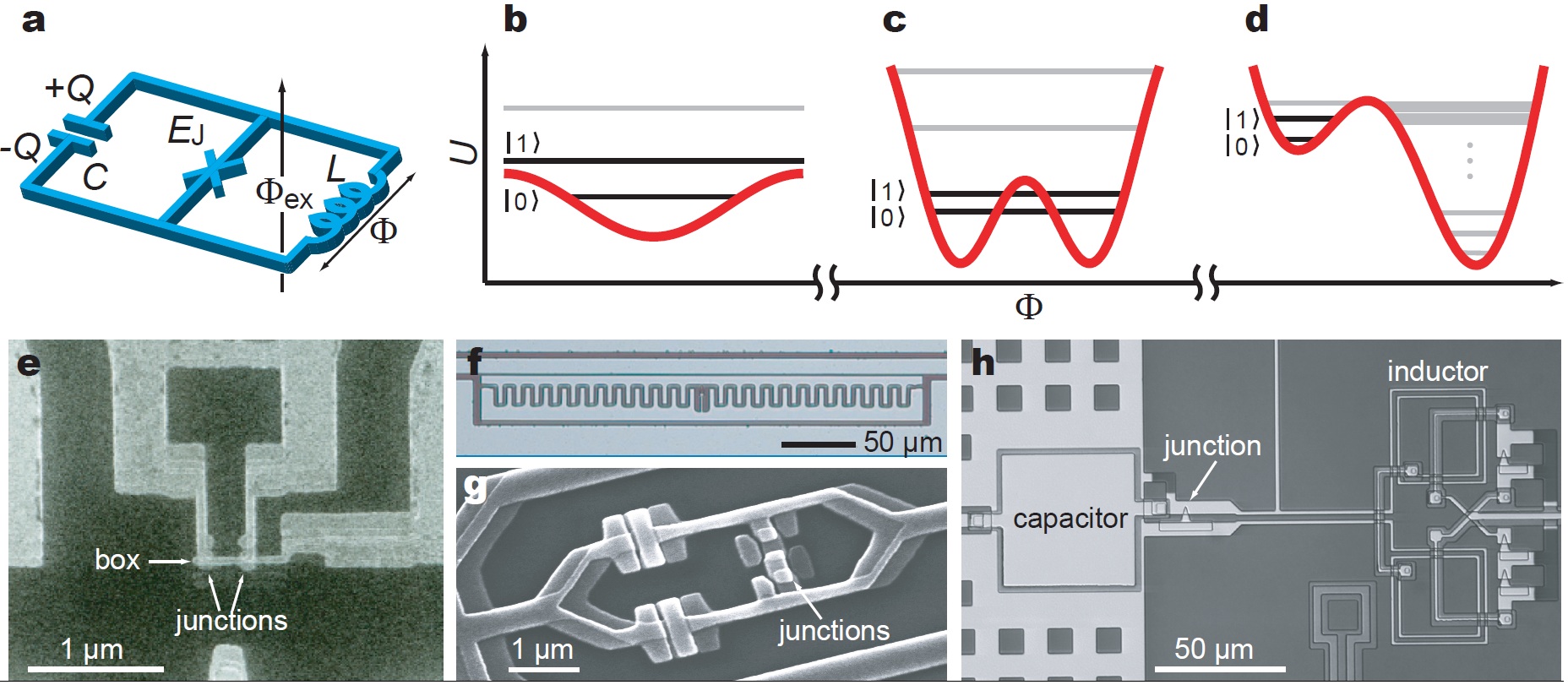}
    \caption\thiscaption\label{superconducting_qubits}
}
{
  \begin{tabular}{m{0.66\textwidth}m{0.34\textwidth}}
  \includegraphics[width=0.66\textwidth]{\thisfigure} &
  \caption{\footnotesize\thiscaption}
\label{superconducting_qubits}
  \end{tabular}
}
\label{superconducting_qubits}
\end{figure*}


All superconducting qubits are realized in electric circuits,
in which one may tune the potential and therefore the wave
function by changing the macroscopically fabricated inductance,
capacitance, and the barrier configuration of the qubits.
Likewise, this potential may be dynamically altered by various
means to give complete quantum control. Typically, the qubit
excitation frequency is designed at 5--10~GHz, which is high
enough to avoid thermal population at the low temperatures
available in dilution refrigerators ($\sim 10$~mK; $k_B T/h
\sim 0.2$~GHz) and low enough for ease of microwave
engineering.  The single-qubit gates are implemented with a
resonant microwave pulse of 1--10~ns inducing Rabi
oscillations. Such pulses are delivered to the qubit locally
using on-chip wires.


Thanks to their macroscopic nature, it is straightforward to
couple superconducting qubits to each other; neighboring qubits
couple strongly either capacitively
or inductively.
These direct couplings have allowed simple quantum logic gates
\cite{ypant03,%
      McDermott05,
      plantenberg07},   
and well-controlled generation of entangled states studied by
quantum state tomography\cite{steffen06}.  
However, for large-scale quantum computer architectures, more adjustable
coupling schemes are desirable. Indirect couplings mediated by
a tunable coupler have been developed for switching on and off
the interaction between qubits
\cite{clarkecoupling06,%
      Niskanen07}.
The application of such tunably coupled qubits to adiabatic
quantum computing is also under investigation
\cite{fluxcoupler07,%
      Harris07,%
      Harris09}.


Exchange of quantum information between arbitrary pairs of
distant qubits may be possible by using a quantum bus, or
\emph{qubus}.  Coupling between superconducting qubits and a
microwave transmission-line resonator is a powerful tool for
this purpose.
The one-dimensional resonators have an extremely small
mode volume and thus strong cooperativity factor\cite{jjStrongCoupling04} 
(see Sec.~\ref{sec:cqed}). Qubits can interact via real- or virtual-photon
exchange through the resonant/off-resonant resonator. Such
systems have allowed two-qubit gate operations between qubits
several millimeters apart \cite{
    phasequbitscavity07,chargequbus07,%
    Leek08},  
and also a variety of \CQED-type experiments in the strong
coupling regime
\cite{jjStrongCoupling04,%
      fluxqubus04,%
      jjRabi06,%
      jjnresolve07,%
      jjsinglephotons07,%
      jjlasing07,%
      jjjaynes08,%
      Hofheinz08,
      Hofheinz09}.

The development of coupling schemes with transmission lines and
resonators has opened new and large potentials for quantum          
microwave optics on a chip. Josephson junctions play multiple
roles in these experiments; they are used to create qubits as
artificial atoms, as discussed, but they also act as nonlinear
inductors. The strong qubit-resonator coupling as well as the
strong nonlinearity of resonators involving Josephson junctions
may allow the exploration of unprecedented regimes of quantum
optics, which may, for example, lead to the use of
continuous-variable quantum information in superconducting
circuits.
Still to be demonstrated, for example, are a single microwave
photon detector and on-chip homodyne mixing, which would
further enrich the microwave quantum-optics tool box.


Adding a measuring device to superconducting circuits without
introducing extra decoherence can be challenging.  The
switching behavior of a current-biased Josephson junction at
its critical current is commonly used as a threshold
discriminator of the two qubit states. Such schemes have been
successfully used in many experiments
\cite{mnau02,  %
      vacjpued02,  %
      cnhm03}  %
and achieved a high measurement fidelity above 90\%
\cite{lucero08}, though the qubit state after the readout is
randomized due to measurement back-action.  A recent, promising
development is the demonstration of \QND\ measurements in which
a qubit provides a state-dependent phase shift for an
electromagnetic wave in a transmission line%
\cite{
wsbfmdgs05,siddiqi06,mooijqnd07}. This shift is then read out
by electronics far from the qubit itself, projecting the qubit
into the eigenstate corresponding to the measurement result.
Again, nearly 90\% fidelity has been demonstrated with non-demolition
properties
\cite{
mooijqnd07}.
Highly efficient amplifiers are crucial for further improvement
of the measurement fidelity.
Integrations of quantum-limited amplifiers employing Josephson
junctions may bring huge impact in this direction
\cite{siddiqi05,%
      CastellanosBeltran08}.

%
For effective fault-tolerant quantum computing, it is important
to rapidly initialize qubits. \QND\ measurements followed by
feedback operations may enable this.  Rapid cooling of qubits
may also be induced by microwaves
\cite{jjcooling06,%
      Grajcar08}.


A notable feature of superconducting qubits is their
macroscopic scale: they involve the collective motion of a
large number ($\sim$10$^{10}$) of conduction electrons in
devices as large as 100~$\mu$m.  Common wisdom is that
superpositions of these larger, more \qu{macroscopic} states
should suffer faster decoherence than more \qu{microscopic}
systems, and indeed superconducting qubits have typically had
the fastest decoherence times of all qubits under widespread
development. However, the distressingly short decoherence times
of a few nanoseconds observed in the earliest experiments have
recently been extended to the range of many microseconds.                     
The enhancement was accomplished by improved circuit designs to
make the qubits more robust%
\cite{vacjpued02,
      Bertet05,
      schreier08}, 
by decoupling from the environment%
\cite{Houck08}, 
and by reducing the noise processes that contribute to
decoherence
\cite{Martinis05}. 
Much current work in superconducting circuit development deals
with understanding and eliminating the noise still remaining.
These noise processes vary for each qubit, but often seem to be
connected to microscopic origins such as charge traps and spins
in the amorphous oxides at the tunnel barriers and at the metal
surfaces, or in the dielectrics for the insulating layers of
capacitances and substrates
\cite{Martinis05,
      Sendelbach08}.  
This kind of process is common to multiple solid-state
implementations of qubits; for example, phosphorous in silicon
suffers a similar problem from the SiO$_2$ barrier, even though
SiO$_2$ provides the \qu{cleanest} insulating layer among
semiconductors. Intensive material engineering research may
eventually solve these problems.



Superconducting qubits provide a wide variety of promising
tools for quantum state manipulations in electric circuits.
Beautiful demonstrations of two-qubit quantum algorithms
(Deutsch-Jozsa and Grover search) were reported
recently\cite{DiCarlo09}. With careful engineering, the
fidelities for control and readout will be increased further.
As the observed decoherence rates improve, these tools will
allow more and more complex circuits, providing an optimistic
future for large-scale quantum computation.

\comment{
\\
\\
Other recent interesting experimental results which have not been included:
\begin{itemize}
\item Tavis-Cummings model, Fink et al. (ETH)
\item decay of Fock states, Wang et al. (UCSB)
\item Lamb shift, Fragner et al. (ETH)
\item Autler-Towns splitting, Baur et al. (ETH)
\item non-linear response of vacuum Rabi resonance, Bishop et al. (Yale)
\item geometric phase, Leek et al. (ETH)
\item uncollapse of quantum states, Katz et al. (UCSB)
\end{itemize}

these are nice experiments, but do not fit well in existing
text, which is already long (and complete) enough

}

\section{Other Technologies}
\version{Version 0.5, sketch by Thaddeus, text by Thaddeus and
Fedor, 5/15/09\\
Version 1.0, text by Thaddeus, references incomplete}

The technologies we have discussed for implementing quantum
computers are by no means the only routes under consideration.
A large number of other technologies exhibiting quantum
coherence have been proposed and tested for quantum computers.

As one example, the single photons in photonic quantum
computers could be replaced by single, ballistic electrons in
low-temperature semiconductor nanostructures, which may offer
advantages in the availability of nonlinearities for
interations and in detection.  As another emerging example,
quantum computers based on ions and atoms may benefit from
using small, polar molecules instead of single atoms, as the
rotational degrees of freedom of molecules offer more
possibilities for coherent
control\cite{demille02,Zoller_polar}.

New materials beyond those we have discussed are also being
investigated in the context of quantum computing.  For example,
some researchers continue to search for new systems that
display the positive optical features of self-assembled quantum
dots and diamond \NV\ centres discussed above (atom-like
behavior, semiconductor host, large oscillator strength) while
exhibiting better homogeneity and coherence than quantum dots
and easier routes to integration than diamond. Shallow,
substitutional semiconductor impurities, for example, exhibit
sharp optical bound states near the bandgap and have the
advantages of being substantially more homogeneous and
potentially easier to place with atomic-scale fabrication
techniques, as in the example of phosphorous in silicon.  The
fluorine impurity in ZnSe is one impurity with a similar
binding energy to phosphorous in silicon and a comparable
possibility for isotopic depletion of nuclear spins from the
substrate.  Unlike in silicon, the direct, wide bandgap of ZnSe
affords it an oscillator strength comparable to a quantum dot.
Further, the \abbrev{II-VI} semiconductor system allows
\MBE-based semiconductor alloying techniques not currently
available in diamond. The electron bound to F:ZnSe and the
$^{19}$F nuclear spin may therefore provide excellent optically
controlled qubits; already it has shown promise as a scalable
single photon source\cite{znse_preprint}.

Another system under investigation for optically controlled,
solid-state quantum computation is provided by rare earth ions
in crystalline hosts.  These systems have been known for many
years to show long coherence times for their hyperfine states.
Unfortunately, these impurity ions usually have weak optical
transitions and and therefore cannot be detected at the single
atom level like quantum dots, \NV\ centres in diamond, or
fluorine impurities in ZnSe.  Therefore, like \NMR\ quantum
computing, this approach employs an ensemble. Isolating the
degrees of freedom to define qubits in this ensemble benefits
from the large inhomogeneous broadening of the system, caused
by shifts of the optical transitions of the impurities due to
imperfections of the crystalline host. Remarkably, these static
shifts only weakly affect the width of transition of individual
ions, which may have optical coherence times of milliseconds.
The extremely high ratio of homogeneous to inhomogeneous
broadening (typically 1~kHz vs. 10~GHz for Eu doped YAlO$_3$)
potentially allows the realization of up to $10^7$ readout
channels in the inhomogeneous ensemble.  Qubits can be defined
as groups of ions having a well defined optical transition
frequency, isolated by a narrow bandwidth laser. Unlike in the
case of liquid state \NMR\ quantum registers, the initial state
of rare-earth qubits can be initialized via optical pumping of
hyperfine sublevels of the ground state.

This system has recently seen a demonstration of single-qubit
state tomography\cite{rare_earth_1,rare_earth_2}. Multi-qubit
gates are also possible via the large permanent dipole moment
in both ground and excited electronic states. Very long
coherence times of the ground state also enable the use of rare
earth qubits as an efficient interface between flying and
matter qubits\cite{rare_earth_3,rare_earth_4,rare_earth_5} with
unprecedented storage times for photons up to 10
sec\cite{rare_earth_6}, which is many orders of magnitude
longer than achieved for atomic systems.

Other materials for hosting single-electron-based qubits are
also under consideration.  The carbon-based nanomaterials of
fullerenes\cite{fullerene_proposal,morton_fullerenes},
nanotubes\cite{marcus_nanotube}, and
graphene\cite{loss_graphene} have excellent properties for
hosting arrays of electron-based qubits. Electrons for quantum
computing may also be held in a low-decoherence environment on
the surface of liquid helium\cite{PD99}. Another spin-based
aproach is the use of molecular magnets. Although these
molecules contain many atoms and many electrons, their magnetic
degrees of freedom at low temperature behave as a single
quantum particle, but with a much stronger and therefore
easier-to-measure magnetic moment\cite{Loss_molecular_magnets}.

A further category of exploration for quantum computation is
new methods to mediate quantum logic between qubits, often of
existing types.  A key example of this is the use of
superconducting transmission line cavities and resonators for
qubits other than those based on Josephson junctions, such as
ions\cite{trbz04}, polar molecules\cite{demille_hybrid} and
quantum dots\cite{superconductor_dots}. Edge-currents in
quantum-hall systems present another type of coherent current
which may be useful for wiring quantum computers\cite{pvk98}.
In fact, nearly every type of bosonic field has been explored
for quantum wiring, including lattice phonons in
semiconductors\cite{Li_in_silicon}, phonons in micromechanical
oscillators\cite{ion_nanomechanical}, free
excitons\cite{pcss02} or hybridizations between excitons and
cavity photons in semiconductors\cite{polariton_mediate}, and
spin-waves in magnetic crystals\cite{kow01}. Other ideas in
this category include surface-acoustic waves for shuttling spin
qubits\cite{BSR2000} and plasmonic technologies for shuttling
photonic qubits at sub-wavelength scales\cite{chang_plasmons}.

Other areas of diverse development in quantum computation are
novel means for measurement.  Ultra-sensitive magnetic field
detection techniques with {\AA}ngstrom-resolution such as
magnetic resonance force microscopy (\abbrev{MRFM}) and
spin-dependent scanning-tunneling microscopy (\abbrev{STM}) may
play a role in future quantum computers.  In the other
direction, technologies developed for qubits such as the \NV\
centre in diamond are finding new roles as magnetic field
sensors in diverse
applications\cite{lukin_NV_magnetometer,wrachtrup_NV_magnetometer}.

A final development in quantum computation deserving of mention
here is the use of topologically defined quantum gates to
preserve quantum information.  Such concepts are used to define
fault-tolerant \QEC\ schemes among ordinary qubits\cite{rh07},
but have also been proposed as a method of physical computation
should a physical system be found to implement them.  For
example, a type of quantum excitation with fractional quantum
statistics known as the anyon has been predicted to play a role
in condensed matter systems (in particular, certain aspects of
the fractional quantum-Hall effect).  Theoretical ideas in
implementing quantum logic by the topological braiding of such
particles may offer more advanced future routes to robust
quantum computation\cite{anyon_review}.

\section{Outlook}
\version{Version 0.5, written by Jeremy, added 1/6/09\\
         Version 1.0, added by Ray 6/8, revised by Thaddeus}

In the last 15 years we have discovered that quantum
information is fundamentally more powerful than classical
information, challenging the tenets of computer science. We
have also learned that it is possible, in principle at least,
to quantum compute reliably in the presence of the imperfection
of real devices.  As demonstrated in this article, we have
learned that we do indeed have enough control today to
implement rudimentary quantum algorithms. These elements form
the foundation of a new kind of science and technology based on
those quantum properties of nature that have no classical
analog.

The challenge for the years to come will be to go from
proof-of-principle demonstrations to the engineering of devices
based on quantum principles that are actually more powerful,
more efficient or less costly than their classical
counterparts.  A quantum computer is perhaps the most ambitious
goal of this new science, and it will probably require a few
more decades to come to fruition.  On the way to this goal,
however, we will grow accustomed to controlling the
counterintuitive properties of quantum mechanics, and we will
develop new materials and make new types of sensors and other
technologies.  As we proceed, we will tame the quantum world
and become inured with a new form of technological reality.

\comment{
Quantum computations (regardless of physical realization) have
traditionally been formulated using the quantum circuit model,
a generalization of the circuit model for Boolean logic: qubits
are represented by wires propagating in time from left to
right, subjected to a sequence of quantum logic gates, and
finally measured\cite{ncbook}. In 2001 a remarkable alternative
was proposed in which the computation starts with a particular
massively entangled state of many qubits---a cluster
state---and the computation proceeds via a sequence of single
qubit measurements from left to right that ultimately leave the
rightmost column of qubits in the answer state \cite{rb01}.
Many physical realizations stand to benefit enormously from
adopting this approach.

A critical consideration for all realizations, is fault
tolerance\cite{ncbook}. In contrast to conventional computers,
quantum computers will be very susceptible to noise, which must
be encoded against (in addition to the encoding described
above). The threshold theorem says that if the noise is below
some threshold an arbitrarily long quantum computation can be
realized. One of the most encouraging results for all
approaches to quantum computing was the high threshold of 1\%
recently reported by Knill\cite{knill05}. Because cluster state
approaches do not conform to the standard model the threshold
theorem does not apply; fortunately analogous thresholds have
been shown to exist \cite{ni-pra-71-042323}. Realistic error
models couple with schemes that are tailored to the specific
physical realization are a major challenge (threshold results
only currently exist for ion trap and optical implementations),
but are essential to the future of quantum computing.

It is clear that the pursuit of a large scale quantum computer
is one of the grandest challenges humankind has ever embarked
upon. Already quantum cryptography systems are now commercially
available, and the next generation of quantum networks will
required quantum repeaters or relays to realize a truly quantum
network and extend their range. This will likely involved
small-scale quantum processors acting on both flying (photonic)
and stationary qubits. Many now believe that a quantum
simulator with hundreds of qubits will ultimately be realized
and could be used to simulate important condensed matter
physics systems and possibly technologically relevant
materials. Whether a large scale factoring machine with
millions of qubits will ever be realized is still an open
question. However, it is clear that along the way we will gain
unprecedented control over quantum systems and more than likely
be surprised by some fundamental science.}


\end{document}